\documentclass[aps,showpacs,showkeys,amsmath,amssymb,preprint,prd,%
floatfix,superscriptaddress]{revtex4}

\usepackage{graphicx}
\usepackage{dcolumn}
\usepackage{docs}%
\usepackage{bm}
\begin{document}
\title{Flavor Diagonal and Off-Diagonal Susceptibilities 
 in a Quasiparticle Model of the Quark-Gluon Plasma}

\author{M.~Bluhm}
\affiliation{
Forschungszentrum Dresden-Rossendorf, PF 510119, 01314 Dresden, Germany}
\author{B.~K\"ampfer}
\affiliation{
Forschungszentrum Dresden-Rossendorf, PF 510119, 01314 Dresden, Germany}
\affiliation{Institut f\"ur Theoretische Physik, TU Dresden, 
01062 Dresden, Germany}

\date{\today}


\begin{abstract}
 The Taylor coefficients of flavor diagonal and off-diagonal susceptibilities 
 as well as baryon number, isovector and electric charge susceptibilities 
 are considered within a
 phenomenological quasiparticle model of the quark-gluon plasma
 and successfully compared with available lattice QCD data 
 up to fourth-order for two degenerate quark flavors. 
 These susceptibility coefficients represent sensible 
 probes of baryon density effects in the equation of state. The baryon 
 charge is carried, in our model, by quark-quasiparticle
 excitations for hard momenta. 
\end{abstract}

\maketitle

\section{Introduction \label{sec:intro}}

The last few years witnessed two important milestones in the realm
of relativistic heavy-ion collisions and related applied QCD: 
(i) Hints for a strongly coupled quark-gluon plasma have been deduced
from experiments at RHIC~\cite{RHIC_announcement,McLerran,Teaney}, and 
(ii) lattice QCD calculations have been extended to non-zero net baryon density.
While for many observables, at mid-rapidity, baryon-density effects are fairly
small in heavy-ion collisions at top RHIC and future LHC energies, 
they become important for the ongoing low-energy runs at RHIC, previous CERN-SPS 
and future FAIR energies. Furthermore, the debated QCD 
critical point seems to be located at non-zero net baryon 
density according to investigations reported in~\cite{CEP}. 
Therefore, the exploration of this part of the
phase diagram of strongly interacting matter 
gains increasing attention, both experimentally and
theoretically. A necessary prerequisite in the search of the
critical point is the understanding of thermodynamic bulk properties of QCD matter 
at non-zero net baryon density.

First-principle lattice QCD calculations include all features of the
complexity of QCD at finite temperature and net baryon density, supposed
the numerical accuracy is appropriate. Indeed, signals of the QCD 
critical point have been found~\cite{Fodor,GavaiCEP}, and the 
pseudo-critical curve not too far from the temperature axis
in the $T - \mu$ plane is routineously determined 
today~\cite{Fodor,Allton} ($T$ and $\mu$ denote
temperature and chemical potential, respectively).

Basically, the partition function $Z(T,\mu)$ or the grand canonical potential
$\Omega(T,\mu)$ or the pressure $p(T,\mu)$ contain much information on
thermodynamic bulk properties of a medium. Susceptibilities are 
second-order derivatives of the pressure in the chemical potential direction. In
so far, susceptibilities represent sensible quantities probing 
the active baryonic degrees of freedom in a medium. Even more,
susceptibilities are related to fluctuations, which are debated to represent 
signatures of deconfinement effects~\cite{Koch,Gavaisigns1}, thus being
of utmost experimental relevance.

Information on various susceptibilities have been accumulated from 
first-principle lattice QCD 
calculations~\cite{GavaiCEP,Gavai,Allton1,Allton2,Gavai1,Maezawa,Hietanen,Bernard,Mukherjee}. 
Keeping in mind possible limitations due
to finite-size, numerical set-up and quark mass ($m_i$) effects they, 
nevertheless, are a source of important information on baryon density effects 
in the hot quark-gluon medium. Below $T_c$, where $T_c$ denotes the
pseudo-critical temperature for deconfinement, 
the hadron resonance gas model (cf.~\cite{Allton2,Karsch}) has been
successfully compared with the lattice QCD data~\cite{Allton2}. Above
$T_c$, the situation is less settled. Certain baryonic bound states
have been considered in~\cite{Shuryak} aiming at arriving at a
physical picture of the strongly coupled quark-gluon plasma. 
Further developments~\cite{Shuryak-Zahed,Peshier-Cassing}
try to deduce also transport properties of deconfined strongly interacting matter. 
In addition, susceptibilities have been studied in phenomenological approaches 
such as the Nambu--Jona-Lasinio 
model~\cite{Sasaki1} or Polyakov loop extensions 
thereof~\cite{Sasaki2,Ghosh,Roessner} as well as quasiparticle 
models~\cite{Bannur,Bluhm_PoS}. Furthermore, within the 
{\boldmath$\Phi$} functional approach to QCD~\cite{Blaizotchi}, 
qualitative agreement with the lattice QCD data in~\cite{Gavai} was found for 
$T\ge 1.5\,T_c$. 
All these approaches attempt to catch the relevant
excitation modes. One should keep in mind that in the weak-coupling
regime or in a medium with prominent quasiparticle excitations, the bulk
properties are governed by excitations with hard momenta 
$k \sim T,\mu$. Soft or ultra-hard modes are expected to influence $p(T,\mu)$ rather less.

For the ultimate description of the very nature of the quark-gluon
plasma one has to know correlations and spectral functions,
propagators and related dispersion relations. Such information is
still fairly scarce, but starts accumulating~\cite{Petretsky,Karsch-Kitazawa}. 
Having at our disposal only the numerical data of thermodynamic
state variables one must try to figure out which physical picture(s)
is (are) compatible. Such an endeavor is clearly phenomenological.
Besides the motivation of getting an interpretation of the available lattice
QCD data, also the applicability of such phenomenological models represents 
important aspects, say for extrapolations to larger net baryon density, or 
for interpolations between different regions of the QCD equation of 
state (EoS)~\cite{Bluhm_PoS}, or for comparing different 
flavor numbers. 

The aim of the present paper is to confront in some detail the lattice
QCD data from~\cite{GavaiCEP,Allton2,Gavai1} for two degenerate 
quark flavors with our quasiparticle 
model~\cite{Peshier1,Peshier2,Bluhm_PLB}. 
This model was used up to now for one
independent chemical potential. With the goal of analyzing isovector, 
electric charge and flavor (off-)diagonal susceptibilities we are going to generalize the
model towards two independent chemical potentials $\mu_u$ and $\mu_d$ 
of up and down quarks, respectively. In fact, isovector and flavor (off-)diagonal
susceptibilities represent much more sensible tests of a model than the 
baryon susceptibility alone. Furthermore, a detailed knowledge about the 
dependence of bulk thermodynamic quantities on, at least, two 
separate quark chemical potentials $\mu_u$ and $\mu_d$ is necessary 
in order to discuss the impact of changes in various flavor sectors 
on the baryon density dependence of the EoS. Also, 
this becomes important when discussing properties of deconfined 
quark matter such as $\beta$-stability and electric charge neutrality in 
hypothetical ultra-dense and hot proto-neutron stars. While 
for one independent chemical potential, say $\mu_u=\mu_d$, 
the model has been successfully 
compared with various sets of lattice QCD data 
in~\cite{Bluhm_PLB,Bluhm_PRC}, the straightforward generalization to a 
set of chemical potentials $\vec{\mu}=\{\mu_u,\mu_d\}$ is restricted by 
consistency requirements~\cite{Peshier_priv} given by Maxwell type relations 
and the stationarity condition of the thermodynamic 
potential. We show here, that the Taylor expansion 
coefficients of various susceptibilities are accessible 
up to a certain order in a consistent formulation, 
contrasting our model as an alternative to the picture 
developed, for instance, in \cite{Shuryak}. 

Our paper is organized as follows. In section~\ref{sec:QPM}, we extend the previous
quasiparticle model \cite{Peshier1,Peshier2,Bluhm_PLB} towards including two
independent chemical potentials and discuss the consistency conditions
for the resulting generalized system of flow equations. Section~\ref{sec:numerics} is devoted to
the numerical evaluation of various susceptibilities and the comparison 
with lattice QCD data. In addition, these results are used for discussing 
some properties of hot deconfined quark matter by means of a Taylor 
expansion of bulk thermodynamic quantities imposing, for instance, 
$\beta$-equilibrium and electric charge neutrality. 
The summary and discussion can be found in section~\ref{sec:conclusions}. 
Appendix A summarizes the entropy density expression and its relation
to the primary thermodynamic potential, while explicit 
representations of coefficients needed for determining the
susceptibilities are listed in Appendices B, C and D. 

\section{Extending the quasiparticle model \label{sec:QPM}}

We consider the case of two degenerate quark flavors for which the 
lattice QCD data~\cite{GavaiCEP,Allton2,Gavai1} are at our disposal. 
We choose the pressure $p(T,\mu_u,\mu_d) = \frac{T}{V} \ln Z(T,\mu_u,\mu_d)$ 
as fundamental quantity in the following 
with quark flavor chemical potentials $\mu_{u,d}$ or, equivalently, 
quark and isovector chemical potentials $\mu_{q,I}$ which 
are related via 
$\mu_q = \frac12 (\mu_u + \mu_d)$, $\mu_I = \frac12 (\mu_u - \mu_d)$ or 
$\mu_u=\mu_q+\mu_I$, $\mu_d=\mu_q-\mu_I$. 
(Note that $\mu_I$ was defined differently as either $\mu_I=2 (\mu_u-\mu_d)$ 
in~\cite{Allton1} or $\mu_I=\frac14 (\mu_u-\mu_d)$ in~\cite{GavaiCEP,Gavai1}.) 
$\mu_q=\frac13 \mu_B$ and $\mu_I$ are 
associated with conserved quantum numbers (in strong interaction 
processes) of baryon number and 
isospin, respectively. The explicit model expression of $p(T,\mu_u,\mu_d)$ 
is relegated to Appendix A. 
The generalized quark number susceptibilities are defined by
\begin{equation}
\label{equ:coeff1}
\chi_{j_u,j_d}(T)=\left.
\frac{\partial^{(j_u + j_d)} p(T,\mu_u,\mu_d)}{\partial\mu_u^{j_u} \, \partial\mu_d^{j_d}}
\right|_{\mu_u=\mu_d=0}.
\end{equation}
Because $\ln Z(T,\mu_u,\mu_d)$ is symmetric under CP transformations, 
derivatives for odd $(j_u + j_d)$ vanish. Furthermore, 
in the flavor symmetric case $m_u=m_d=m$, we find 
$\chi_{j_u,j_d}(T)=\chi_{j_d,j_u}(T)$. 

These generalized quark number susceptibilities represent a rich test ground.
Besides the mentioned physical meaning of susceptibilities as measures for fluctuations,
they additionally constitute the Taylor coefficients of the excess pressure
$\Delta p(T,\mu_u,\mu_d) \equiv p(T,\mu_u,\mu_d) - p(T,\mu_u=0,\mu_d=0)$, 
expanded simultaneously in powers of $\mu_u$ and $\mu_d$ via 
\begin{equation}
\label{equ:excess}
\Delta p(T,\mu_u,\mu_d)=\sum_{j_u,j_d} 
\chi_{j_u,j_d}(T) \frac{\mu_u^{j_u}}{j_u!} \frac{\mu_d^{j_d}}{j_d!} \,,
\end{equation}
thus containing information about baryon density effects in the EoS. 
(Similarly, one could consider the Taylor expansion of the pressure 
$p(T,\mu_q+\mu_I,\mu_q-\mu_I)$ in terms of $\mu_q$ and $\mu_I$.)
Various associated expansion coefficients, e.~g.~those of flavor diagonal 
and off-diagonal susceptibilities as well as relations among them, 
are discussed below. The excess pressure becomes increasingly important 
in the domain of larger values of $\mu_{u,d}$ and lower temperatures. 
For instance, imposing $\beta$-stability and electric charge neutrality 
of hot quark matter stars requires the knowledge about the dependence of bulk 
thermodynamic quantities on $\mu_u$ and $\mu_d$ separately, 
at least. This underscores the importance of the susceptibilities 
$\chi_{j_u,j_d}(T)$ even if, at small $T$, a Taylor expansion 
in $\mu_{u,d}$ directions may not suffice. 

The net quark flavor number densities 
$n_i = \partial p / \partial \mu_i$ with $i = u,d$ read 
\begin{eqnarray}
\label{equ:QPM9}
n_i & = & \frac{d_i}{2\pi^2} \int_0^\infty dk k^2 
\left[\frac{1}{e^{(\omega_i-\mu_i)/T}+1} - 
\frac{1}{e^{(\omega_i+\mu_i)/T}+1}\right]\,, 
\label{equ:QPM10}
\end{eqnarray}
where $d_i = d_{u,d} = 2\,N_c$ refers to the spin and color degeneracies of the quarks. 
This implies that baryon charge-$\frac 13$ carriers are quasiparticles with quark quantum numbers
obeying the dispersion relations~\cite{Seipt_Diploma} 
\begin{equation}
\label{eq.5}
\omega^2_i = k^2 + m_i^2 + \Pi_i, \quad 
\Pi_i = \frac13 \left(T^2 + \frac{\mu_i^2}{\pi^2} \right) G^2 (T,\mu_u,\mu_d)\,. 
\end{equation}
For later purposes, we also exhibit 
the corresponding expression for gluons reading 
\begin{equation}
\omega^2_g = k^2 +         \Pi_g, \quad 
\Pi_g = \frac23 \left(T^2 + \frac{3}{8 \pi^2}(\mu_u^2 + \mu_d^2) \right)G^2 (T,\mu_u,\mu_d)\,. 
\label{eq.7}
\end{equation}
The quark mass parameters $m_i$ might comply with the lattice calculational set-up, 
e.g.\ either $m_i = \xi_i T$ with constant $\xi_i$ to compare 
with~\cite{Allton2}, or constant $m_i$ to compare with~\cite{GavaiCEP,Gavai1}. 

The crucial point is that, besides the displayed explicit dependence of the 
self-energy parts $\Pi_l$ (where $l$ is a label for $u, d, g$) on 
$\{\mu_i\} = \mu_{u,d}$ and $T$, 
there is also an implicit dependence
via the effective coupling $G^2(T, \mu_u, \mu_d)$. 
In the case of one independent chemical potential $\mu_u=\mu_d=\mu_q$ (i.~e.~$\mu_I=0$), 
the Maxwell relation $\partial n_q / \partial T = \partial s / \partial \mu_q$, 
with entropy density $s$, 
together with the stationarity of the grand canonical potential,
$\delta p/ \delta \Pi_l = 0$, leads to Peshier's flow 
equation~\cite{Peshier2} which determines $G^2(T,\mu_q)$ for 
given initial condition $G^2(T,\mu_q=0)$. In the case of two independent
chemical potentials $\mu_{u,d}$ or, equivalently, $\mu_{q,I}$, 
a system of three coupled equations is obtained from demanding stationarity 
and from the Maxwell relations 
\begin{eqnarray}
\label{equ:QPM31}
\frac{\partial s}{\partial\mu_I} &=& \frac{\partial n_I}{\partial T} , \\
\label{equ:QPM32}
\frac{\partial s}{\partial\mu_q} &=& \frac{\partial n_q}{\partial T}, \\
\label{equ:QPM33}
\frac{\partial n_I}{\partial\mu_q} &=& \frac{\partial n_q}{\partial\mu_I} \,.
\end{eqnarray} 
The needed expression for the entropy density is listed in Appendix A and, 
with the definitions of $\mu_{u,d}$ and $\mu_{q,I}$ above, 
we note for isovector and quark number densities $n_I=n_u-n_d$ 
and $n_q=3\,n_B=n_u+n_d$, respectively. 
The emerging system generalizes Peshier's flow equation~\cite{Peshier2} 
towards two independent chemical potentials propagating 
$G^2(T,\mu_q=0,\mu_I=0)$ into the thermodynamic parameter space, 
i.~e.~to non-zero $\mu_q$ and $\mu_I$. 

The structure of the generalized system of flow equations reads in the 
basis ($\mu_q,\mu_I$) 
\begin{eqnarray}
\label{eq.11} 
A_1 \frac{\partial G^2}{\partial\mu_I} + B_1 \frac{\partial G^2}{\partial T} & = & C_1 , \\
\label{equ:QPM38} 
A_2 \frac{\partial G^2}{\partial\mu_q} + B_2 \frac{\partial G^2}{\partial T} & = & C_2 , \\
\label{eq.13} 
(A_3-B_3) \frac{\partial G^2}{\partial\mu_q} & = & (A_3+B_3) 
\frac{\partial G^2}{\partial\mu_I} , 
\end{eqnarray}
with coefficients $A_{1,2,3}$, $B_{1,2,3}$, $C_{1,2}$ listed in Appendix B. 
From these coefficients it becomes evident how quark and gluon sectors are coupled. 
It was earlier argued~\cite{Peshier_priv} that the generalized system 
of flow equations in Eqs.~(\ref{eq.11}-\ref{eq.13}) cannot be solved uniquely 
for arbitrary values of $\mu_q$ and $\mu_I$, but only when assuming a side 
condition $\mu_u=\mu_u(\mu_d)$, i.~e.~when considering merely one independent 
chemical potential. To see this, we reformulate 
Eqs.~(\ref{eq.11}-\ref{eq.13}) in the basis ($\mu_u,\mu_d$), making use 
of the analog of Eq.~(\ref{eq.13}) in terms of $\mu_u$ and $\mu_d$. Then, 
the generalized system of flow equations is transformed into 
\begin{eqnarray} 
\label{eq.14}
\mathcal{A}_1 \frac{\partial G^2}{\partial \mu_u} + \mathcal{B}_1 \frac{\partial G^2}{\partial T} & = & 
\mathcal{C}_1 \,, \\
\label{eq.15}
\mathcal{A}_2 \frac{\partial G^2}{\partial \mu_u} + \mathcal{B}_2 \frac{\partial G^2}{\partial T} & = & 
\mathcal{C}_2 \,. 
\end{eqnarray} 
These partial differential equations are uniquely solvable if the coefficients 
${\cal A}_{1,2}$ and ${\cal B}_{1,2}$ and ${\cal C}_{1,2}$, as 
listed in Appendix C, are pairwise equal. 
Indeed, ${\cal A}_{1}={\cal A}_{2}$ 
and ${\cal B}_{1}={\cal B}_{2}$ hold in general. But, the furthermore 
needed equality ${\cal C}_{1}={\cal C}_{2}$ is ensured for arbitrary 
but small values of $\mu_{u,d}$, i.~e.~$\mu_{u,d}\ll\pi T$. (Actually, 
the equality of ${\cal C}_{1}$ and ${\cal C}_{2}$ is given up to 
order $\mathcal{O}(\mu_{u,d}^2)$ in a Taylor series expansion in terms 
of $\mu_u$ and $\mu_d$; the coefficients of third-order terms start 
to differ.) 

As the primary goal of this paper is the comparison of 
susceptibility Taylor series expansion coefficients Eq.~(\ref{equ:coeff1}) 
(to be calculated as derivatives of $p$ at $\mu_{u,d}=0$) 
with lattice QCD results up to fourth-order, the necessary conditions 
are fulfilled. The issue of potential limitations provided by the 
restriction to the small $\mu_{u,d}$ region and one possible way 
of circumventing them are discussed in Appendix~D. In the needed 
leading order for evaluating the susceptibility coefficients 
of interest we note that from Eqs.~(\ref{equ:A8}) and~(\ref{equ:A9}) in 
Appendix~C 
\begin{eqnarray} 
\label{equ:A10}
\mu_u & = & \mu_d\frac{\mathcal{I}_1}{\mathcal{I}_2} \,,\\
\label{equ:A11}
\mu_u & = & \mu_d\frac{\mathcal{I}_5}{\mathcal{I}_4}\frac{\mathcal{I}_1}{\mathcal{I}_2} 
\,,
\end{eqnarray}
where ${\cal I}_{k}$ represent phase-space integrals listed in Appendix~B, 
implying also $\omega_u=\omega_d$ in the 
mass symmetric case, $m_u=m_d$, and ${\cal I}_4={\cal I}_5$. 

Furthermore, by exploiting Eqs.~(\ref{eq.11}-\ref{eq.13}), one finds 
$\left. \frac{\partial G^2}{\partial\mu_q}\right|_{\mu_q=\mu_I=0} = 
\left. \frac{\partial G^2}{\partial\mu_I}\right|_{\mu_q=\mu_I=0} = 0$ 
and 
\begin{eqnarray} 
\label{eq.18}
\left.\frac{\partial^2 G^2}{\partial\mu_q^2}\right|_{\mu_q=\mu_I=0} 
& = & 
\frac{1}{\mathcal{N}} \bigg\{ \mathcal{N}_1 
\left[2\xi_u^2T + \frac23 T G^2(T) + 
\frac13 T^2 \frac{\partial G^2(T)}{\partial T}\right] \\
\nonumber & & + 
\mathcal{N}_2 \left[2\xi_d^2T + \frac23 T G^2(T) + 
\frac13 T^2\frac{\partial G^2(T)}{\partial T}\right] 
\\ 
\nonumber & & - \mathcal{I}_3 \frac{1}{\pi^2} G^2(T) - 
\mathcal{I}_4 \frac{2}{3\pi^2} G^2(T) - 
\mathcal{I}_5 \frac{2}{3\pi^2} G^2(T) \bigg\} \,,
\end{eqnarray}
while from Eqs.~(\ref{eq.14}) and~(\ref{eq.15}), we find 
$\left. \frac{\partial G^2}{\partial\mu_u}\right|_{\mu_u=\mu_d=0} = 
\left. \frac{\partial G^2}{\partial\mu_d}\right|_{\mu_u=\mu_d=0} = 0$ 
and 
\begin{eqnarray} 
\label{eq.18_new}
\left.\frac{\partial^2 G^2}{\partial\mu_u^2}\right|_{\mu_u=\mu_d=0} 
& = & 
\frac{1}{\mathcal{N}} \bigg\{ \mathcal{N}_1 
\left[2\xi_u^2T + \frac23 T G^2(T) + 
\frac13 T^2 \frac{\partial G^2(T)}{\partial T}\right] \\
\nonumber & & - \mathcal{I}_3 \frac{1}{2\pi^2} G^2(T) - 
\mathcal{I}_4 \frac{2}{3\pi^2} G^2(T) \bigg\} \,,
\end{eqnarray}
with coefficients ${\cal N}, \, {\cal N}_{1,2}$ listed in Appendix C, 
$G^2(T)=G^2(T,\mu_u=0,\mu_d=0)$ and ${\cal I}_{3,4,5}$ considered at 
$\mu_q=\mu_I=0$ or, equivalently, $\mu_u=\mu_d=0$. Note that in the flavor 
symmetric case considered here, Eqs.~(\ref{eq.18}) and~(\ref{eq.18_new}) are 
related via $\left.\partial^2G^2/\partial\mu_u^2\right|_{\mu_{u,d}=0} = 
\frac12 \left.\partial^2G^2/\partial\mu_q^2\right|_{\mu_{q,I}=0}$. In 
addition, odd derivatives with respect to the chemical potentials such 
as $\frac{\partial^3 G^2}{\partial\mu_u^3}$ or mixed derivatives such as 
$\frac{\partial^2 G^2}{\partial\mu_u\partial\mu_d}$ or 
$\frac{\partial^2 G^2}{\partial\mu_q\partial\mu_I}$ vanish at 
$\mu_u=\mu_d=0=\mu_q=\mu_I$. The above stated expressions and equalities are 
uniquely obtained from the generalized system of flow equations in 
the region of small $\mu_{u,d}$, and we can proceed by evaluating 
various susceptibilities. 

\section{Comparison with lattice QCD data \label{sec:numerics}}
\subsection{Taylor expansions in {\boldmath$\mu_q/T$} at {\boldmath$\mu_I=0$} \label{sec:3_1}}

In this section, we confront the above extended quasiparticle model (QPM) 
with lattice QCD data of various susceptibilities for $N_f=2$ degenerate 
quark flavors. In~\cite{Allton2}, quark number and isovector 
susceptibilities as well as flavor diagonal and off-diagonal susceptibilities 
have been calculated on a lattice with temporal and spatial 
extensions $N_\tau=4$ and $N_\sigma=16$ using improved actions and 
$m_u=m_d=0.4\,T$ (i.~e.~$\xi_u=\xi_d=0.4$) as quark mass parameters. 
As a special case of the 
Taylor expansion in terms of $\mu_q$ and $\mu_I$, expansions in terms 
of $\mu_q/T$ at $\mu_I=0$ were considered. 

The quark number susceptibility 
$\chi_q(T,\mu_q) / T^2 = \frac{\partial^2 (p(T,\mu_q,\mu_q)/T^4)}
{\partial (\mu_q/T)^2} = 2 c_2 + 12 c_4 \left(\frac{\mu_q}{T}\right)^2 + 
30 c_6\left(\frac{\mu_q}{T}\right)^4 + {\cal O} (\mu_q^6)$ with 
$c_k(T)=\left.\frac{1}{k!}
\frac{\partial^k \left( T^{-4} p (T,\mu_q+\mu_I,\mu_q-\mu_I)\right)}{\partial  (\mu_q/T)^k}\right|_{\mu_{q,I}=0}$ 
has been analyzed already in detail in \cite{Bluhm_PLB}, and an impressively good
agreement of our model with the lattice QCD data from \cite{Allton2} has been found.
The isovector susceptibility $\chi_I(T,\mu_q)$ is only accessible with the present extension
of our model; it obeys the expansion
\begin{equation}
\label{equ:QPM23}
\frac{\chi_I(T,\mu_q)}{T^2} = \frac{\partial^2 (p(T,\mu_q+\mu_I,\mu_q-\mu_I)/T^4)}{\partial (\mu_I/T)^2} = 2c_2^I + 
12c_4^I\left(\frac{\mu_q}{T}\right)^2 + 30c_6^I\left(\frac{\mu_q}{T}\right)^4 + 
{\cal O} (\mu_q^6) \,,
\end{equation}
where the expansion coefficients read 
\begin{equation}
\label{equ:QPM24}
c_k^I(T)=\left.\frac{1}{k!}
\frac{\partial^k \left( T^{-4} p (T,\mu_q+\mu_I,\mu_q-\mu_I)\right)}{\partial (\mu_I/T)^2 
\partial (\mu_q/T)^{k-2}}\right|_{\mu_{q,I}=0}\,. 
\end{equation}
Due to the invariance of $\ln Z$ under CP transformations, $c_k^I$ vanish for odd $k$. 

For the first coefficient of interest we find within the extended QPM 
the explicit representation 
\begin{equation}
\label{equ:QPM27}
c_2^I(T) = \frac{d}{\pi^2} \int_0^\infty dk \frac{k^2}{T^3} 
\frac{e^{\epsilon_{0}}}{(e^{\epsilon_{0}}+1)^2} \,, 
\end{equation}
with $d=d_u=d_d$ and 
$\epsilon_0 = \omega_u/T|_{\mu_{u,d}=0} = \omega_d/T|_{\mu_{u,d}=0}$ 
which implies $c_2^I=c_2$ for all temperatures (cf.~\cite{Bluhm_PLB}) in the QPM. 
The next non-zero coefficient is
\begin{eqnarray}
c_4^I(T) & = & \frac{d}{12\pi^2} \int_0^\infty dk \frac{k^2}{T^3} 
\frac{e^{\epsilon_{0}}}{(e^{\epsilon_{0}}+1)^4} 
\bigg\{ e^{2\epsilon_{0}} - 4e^{\epsilon_{0}} + 1 \\ \nonumber 
& & - 
\frac{(e^{2\epsilon_{0}}-1)}{\epsilon_{0}} 
\left(\frac{1}{\pi^2} G^2(T) + 
\frac{T^2}{6} \left.\frac{\partial^2 G^2}{\partial\mu_q^2}
\right|_{\mu_{q,I}=0} \right) \bigg\}\,, 
\end{eqnarray} 
where $\left.\frac{\partial^2 G^2}{\partial\mu_q^2}
\right|_{\mu_{q,I}=0}$ is given in Eq.~(\ref{eq.18}). 

As in our previous studies \cite{Bluhm_PLB}, we choose for the
effective coupling $G^2(T,\mu_u=0,\mu_d=0)$ entering 
Eqs.~(\ref{eq.5}),~(\ref{eq.7}) and~(\ref{eq.18}) the parametrization 
\begin{eqnarray}
\label{eff_G}
G^2(T) = \left\{
\begin{array}{ll}
G^2_{\rm 2 loop} (\zeta),\hspace*{4mm} \zeta=\lambda \frac{(T-T_s)}{T_c}, &T \ge T_c ,
\\[3mm]
G^2_{\rm 2 loop}(T_c) + b (1 - \frac{T}{T_c}),&T < T_c .
\end{array} 
\right.
\label{eq.20}
\end{eqnarray}
The numerically evaluated QPM results for $c_2^I$ and $c_4^I$ are exhibited in Fig.~1
(results for $c_{2,4}$ are exhibited in~\cite{Bluhm_PLB}) and 
compared with lattice QCD data~\cite{Allton2}, where we use as parameters entering 
$G^2(T)$ in Eq.~(\ref{eff_G}) $\lambda=5.95$, $T_s=0.75\,T_c$ and 
$b=421.5$. 
\begin{figure}[t]
\label{fig.1}
\begin{center}
\includegraphics[scale=0.29,angle=0.]{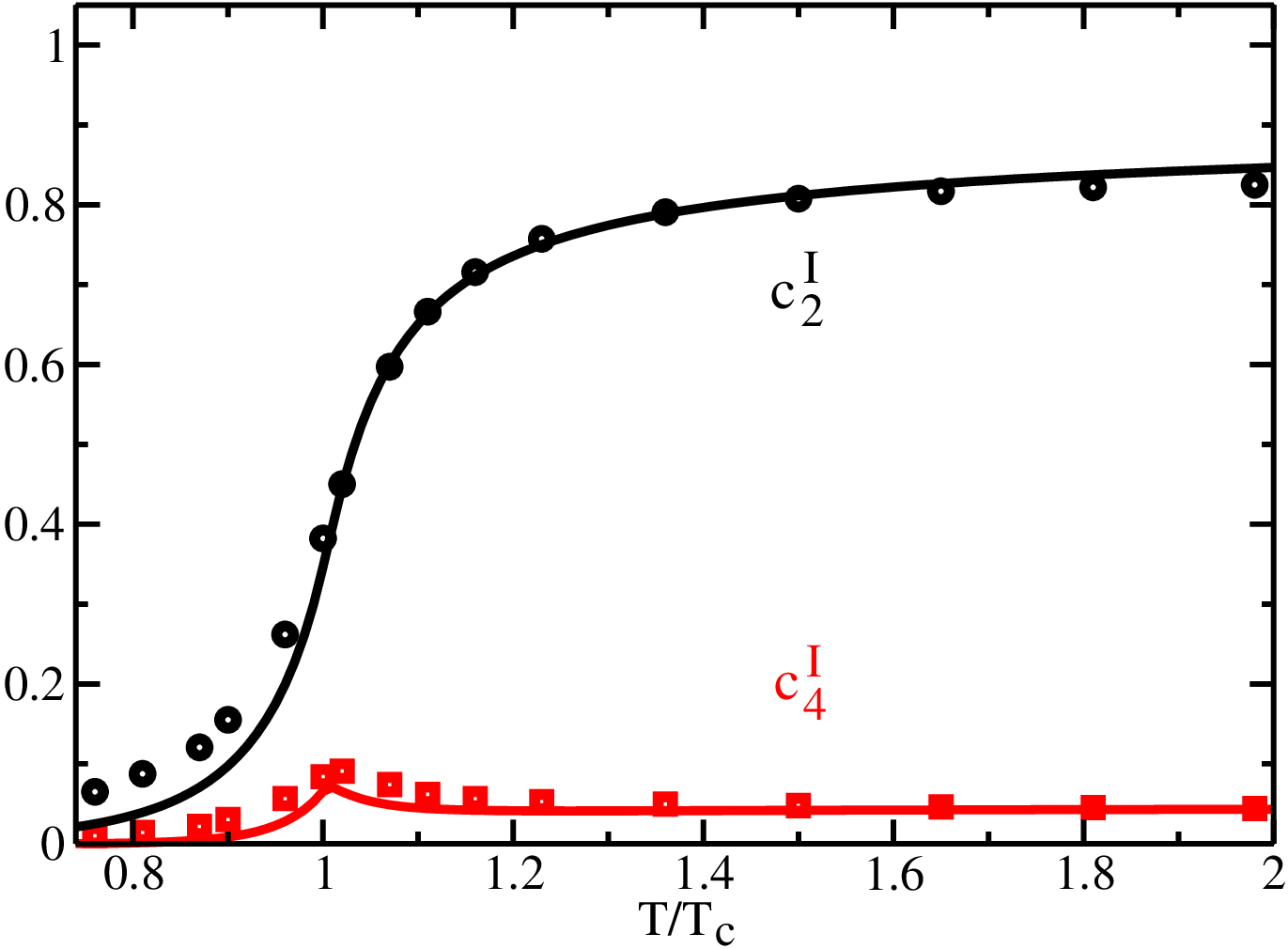}
\caption{(color online). Comparison of QPM results (solid curves) for $c_k^I$ with lattice QCD 
  data~\cite{Allton2} (circles for $k=2$ and squares for $k=4$) for $N_f=2$ quark flavors.}
\end{center}
\end{figure}
The explicit value of $T_c$ is not important for the scaled quantities considered here. 
(Note that these parameters differ from the parametrization 
employed in~\cite{Bluhm_PLB} for describing $c_i$. This 
is due to the different quark dispersion relations used in~\cite{Bluhm_PLB} 
and here. Employing instead Eq.~(\ref{eq.5}) as quark dispersion relation 
with the QPM parameters for $G^2(T)$ stated above, we find an equally 
good agreement of QPM results for $c_i$ with the lattice 
QCD data~\cite{Allton2} as reported in~\cite{Bluhm_PLB}.) 

Similar to $c_4$, the expansion coefficient $c_4^I$ slightly underestimates the 
lattice QCD data \cite{Allton2} approaching its Stefan-Boltzmann limit 
$1/(2\pi^2)$, while $c_2^I$ agrees remarkably well
with the data \cite{Allton2} for $T \ge T_c$ approaching its Stefan-Boltzmann 
limit $N_f/2$ asymptotically. 
Whereas $\chi_q$, due to the growing importance of the higher-order 
expansion coefficients with increasing chemical potential, exhibits a 
significant peak structure close to $T_c$ for large $\mu_q/T$ indicating 
some critical behavior, $\chi_I$ does not point to such structures. 
This behavior is a consequence of the much less pronounced peak 
in $c_4^I$ compared to $c_4$. Similar findings were reported in~\cite{Hatta}, 
where a phenomenological sigma model was considered. 
Below $T_c$, the agreement with lattice 
QCD data might be accidental, but one may consider Eq.~(\ref{eff_G}) as convenient 
parametrization also for this region (see also discussion 
in section~\ref{sec:conclusions}). 

Correlations between fluctuations in different flavor components can be 
discussed by considering flavor diagonal and off-diagonal susceptibilities. 
They read 
\begin{equation} 
\label{equ:QPM51}
\frac{\chi_{uu}}{T^2} = \frac 14 \left(\frac{\chi_q}{T^2}+\frac{\chi_I}{T^2}\right) 
  = 2c_2^{uu} + 12 c_4^{uu}\left(\frac{\mu_q}{T}\right)^2 
  + 30 c_6^{uu}\left(\frac{\mu_q}{T}\right)^4 + {\cal O} (\mu_q^6) 
\end{equation}
for the flavor diagonal susceptibility and 
\begin{equation}
\label{equ:QPM52}
\frac{\chi_{ud}}{T^2} = \frac 14 \left(\frac{\chi_q}{T^2}-\frac{\chi_I}{T^2}\right) 
  = 2c_2^{ud} + 12 c_4^{ud}\left(\frac{\mu_q}{T}\right)^2 
  + 30 c_6^{ud}\left(\frac{\mu_q}{T}\right)^4 + {\cal O} (\mu_q^6) 
\end{equation}
for the flavor off-diagonal susceptibility, where 
the individual expansion coefficients are defined by $c_k^{uu}=(c_k+c_k^I)/4$ and 
$c_k^{ud}=(c_k-c_k^I)/4$. 

The expansion coefficients $c_k^{uu}$ and $c_k^{ud}$ 
for $k=2,4$ are exhibited in Fig.~2 and 
compared with lattice QCD data~\cite{Allton2}. 
\begin{figure}[t]
\label{fig.2}
\includegraphics[scale=0.29,angle=0.]{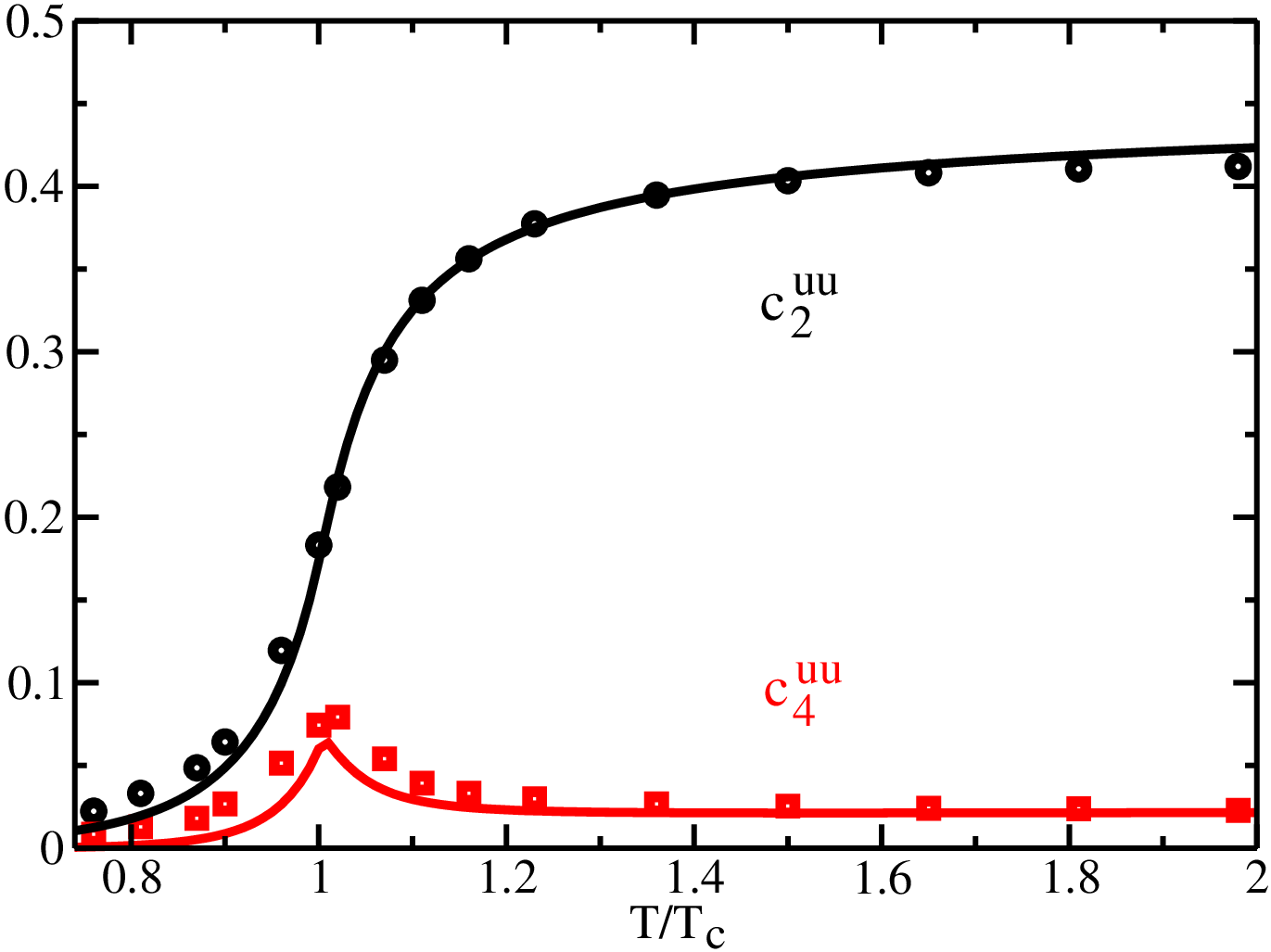}
\hskip 3mm
\includegraphics[scale=0.29,angle=0.]{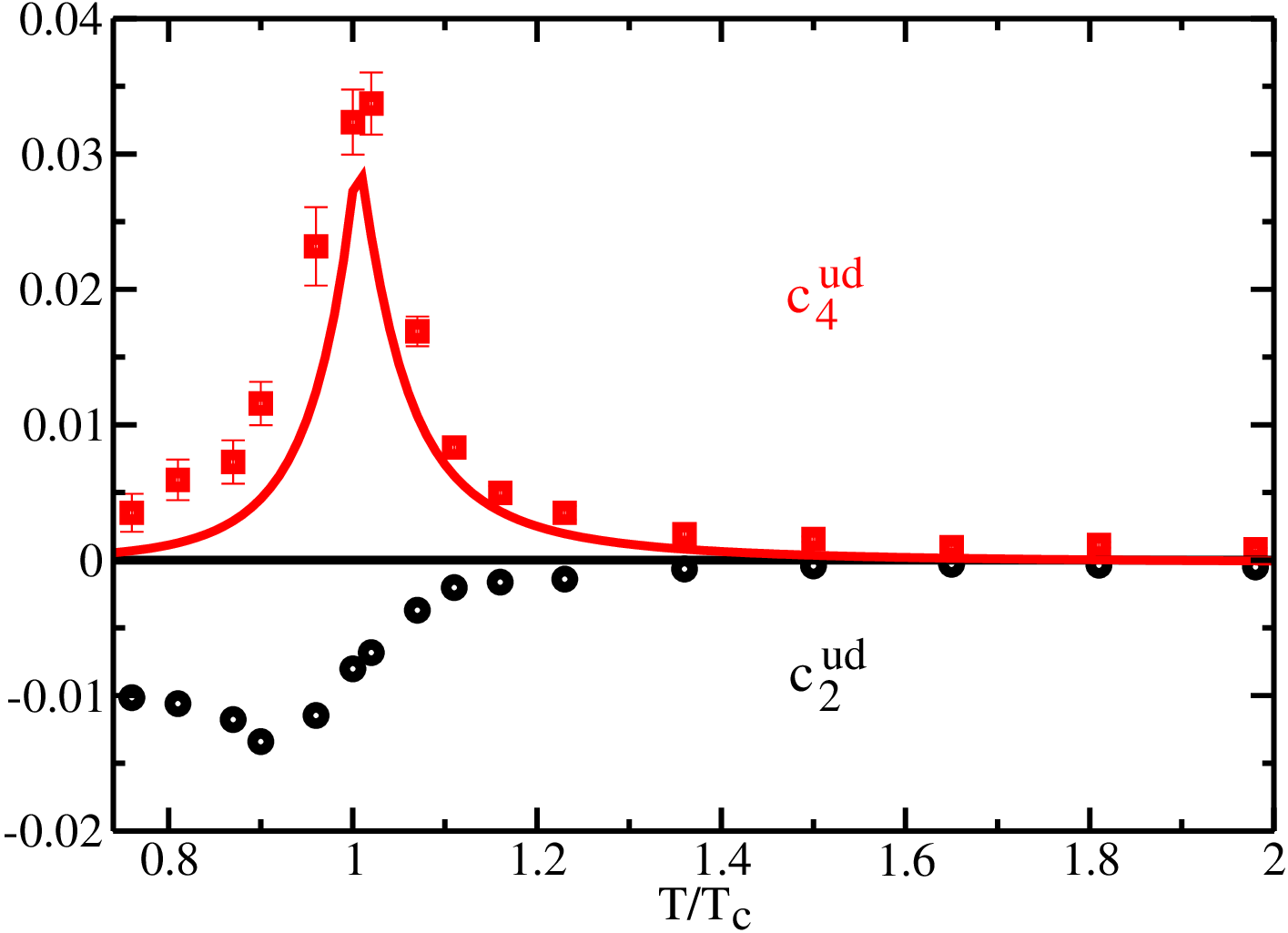}
\caption{(color online). Comparison of QPM results (solid curves) for 
  the expansion coefficients $c_k^{uu}$ 
  of the flavor diagonal susceptibility 
  $\chi_{uu}$ (left) and $c_k^{ud}$ of the flavor off-diagonal susceptibility 
  $\chi_{ud}$ (right) with lattice QCD data~\cite{Allton2} (circles for $k=2$, 
  squares for $k=4$).} 
\end{figure}
The diagonal expansion coefficients $c_{2,4}^{uu}$ show a similar pattern as $c_{2,4}$ and 
$c_{2,4}^I$ approaching their Stefan-Boltzmann limits $N_f/4$ asymptotically 
in the case of $c_2^{uu}$ and $1/(4\pi^2)$ for $T>2\,T_c$ in the case 
of $c_4^{uu}$. The pronounced peak structure of the off-diagonal expansion coefficient
$c_4^{ud}$ is well reproduced, while in our extended QPM $c_2^{ud}$ is zero 
for all temperatures, in contrast to the data 
which are numerically small and differ noticeably from zero 
only in the region $T\lesssim T_c$. This is simply a consequence 
of $c_2^I=c_2$. The pattern observed for the flavor off-diagonal 
susceptibility coefficients is discussed further in 
section~\ref{sec:3_2}. 
As the flavor off-diagonal susceptibility coefficients $c_k^{ud}$ 
rapidly approach their Stefan-Boltzmann limit which is zero for all $k$, 
$\chi_{ud}$ vanishes for large $T$ indicating that fluctuations in different 
flavor channels are uncorrelated at high temperatures. On the other 
hand, $\chi_{ud}$ increases rapidly with increasing $\mu_q$ in the vicinity 
of $T_c$, indicating increasing correlations~\cite{Koch,Gavaisigns1} between fluctuations in 
different flavor channels in the transition region. This also explains 
the observed different behavior of $\chi_q$ and $\chi_I$: While peak structures 
effectively add up in $\chi_q$ they approximately cancel each other in $\chi_I$. 

The behavior of the electric charge susceptibility $\chi_Q$ is strongly 
related to $\chi_q$ and $\chi_I$ via $\chi_Q=\frac 14 \left(\chi_I+\frac 19 \chi_q\right)$. 
The corresponding Taylor expansion reads 
\begin{equation}
\label{equ:chiQ}
\frac{\chi_Q(T,\mu_q)}{T^2} = 2c_2^Q + 
12c_4^Q\left(\frac{\mu_q}{T}\right)^2 + 30c_6^Q\left(\frac{\mu_q}{T}\right)^4 + 
{\cal O} (\mu_q^6) \,,
\end{equation}
with expansion coefficients $c_k^Q=\frac 14 \left(c_k^I+\frac 19 c_k\right) = 
\left(\frac 59 c_k^{uu}-\frac 49 c_k^{ud}\right)$. 
The coefficients $c_2^Q$ and $c_4^Q$ are exhibited in Fig.~3. 
\begin{figure}[t]
\label{fig.3}
\begin{center}
\includegraphics[scale=0.29,angle=0.]{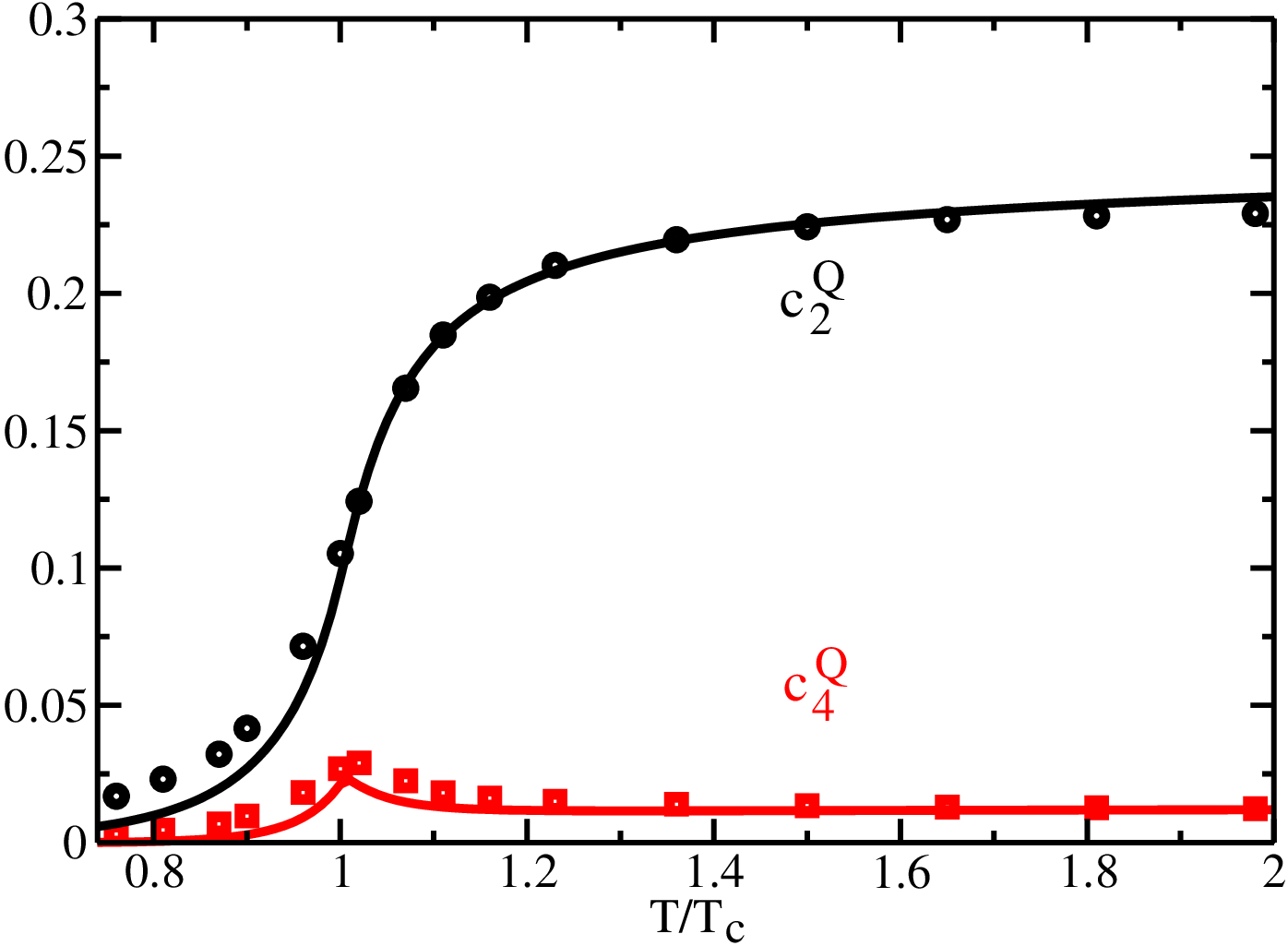}
\caption{(color online). Comparison of QPM results (solid curves) for the electric charge 
susceptibility coefficients $c_k^Q$ with lattice QCD data~\cite{Allton2} 
(circles for $k=2$, squares for $k=4$).}
\end{center}
\end{figure}
From the definition of $c_k^Q$ it becomes clear that contributions of 
pronounced structures appearing in flavor diagonal and off-diagonal susceptibilities 
do not completely cancel in the electric charge susceptibility. 

\subsection{Taylor expansion in {\boldmath$\mu_u$} and {\boldmath$\mu_d$} \label{sec:3_2}}

In this section, we confront the extended QPM with lattice QCD 
data~\cite{GavaiCEP,Gavai1} of 
generalized quark number susceptibilities $\chi_{j_u,j_d}$ as 
defined in Eq.~(\ref{equ:coeff1}). These simulations were also performed 
for $N_f=2$ degenerate quark flavors on a lattice with temporal and 
spatial extensions $N_\tau=4$ and $N_\sigma=16$. However, the used quark 
mass parameter entering the quark dispersion relation 
reads now $m_u=m_d=0.1\,T_c$, in agreement with the lattice 
performance~\cite{GavaiCEP,Gavai1}, which is temperature independent 
in contrast to the lattice set-up considered in section~\ref{sec:3_1}. As a result, 
some of the coefficients in the generalized system of flow equations render which 
changes also the derivative expressions of the effective coupling. To be precise, 
the terms explicitly depending on $\xi_u$ and $\xi_d$, which 
enter Eqs.~(\ref{eq.18}) and~(\ref{eq.18_new}) and some 
coefficients in Appendix B and C, have to vanish for constant $m_{u,d}$. 

Furthermore, non-improved actions have been employed 
in~\cite{GavaiCEP,Gavai1}, thus cut-off effects on the numerical results 
are sizeably increased compared to improved actions. 
In section~\ref{sec:3_1}, we 
assumed the lattice QCD data~\cite{Allton2} to be rather close to the continuum limit 
as improved actions were used (cf. a discussion in~\cite{Karschcutoff}); thus 
no continuum correction factor was applied. (As discussed 
in~\cite{Allton1}, continuum limit corrections to the Taylor expansion 
coefficients $c_k$ are expected to be similar (10-20\%) to corrections 
for the pressure at zero chemical potential~\cite{Karsch1}, even though, the 
corrections seem to increase for higher-order expansion 
coefficients, see~\cite{Allton1,Karschpriv}.) 
Here, however, we have to rely on an 
estimate for the continuum extrapolation of the lattice QCD 
data from~\cite{GavaiCEP,Gavai1}. 
By investigating different temporal lattice extensions $N_\tau$ 
at fixed large temperature 
in~\cite{Gavai2,Gavai3}, the continuum limit of some 
generalized quark number susceptibilities 
was estimated. 
Even though, in principle, correction 
factors could be different for different temperatures, 
we apply as scaling factors $d_{lat}^{(\chi_2)}=0.47$ in the case 
of $\chi_{2,0}/T^2$~\cite{Gavai2,Gavai3} and a larger correction 
$d_{lat}^{(\chi_4)}=0.32$ in the case of $\chi_{4,0}$~\cite{Gavai3} 
to the data~\cite{GavaiCEP,Gavai1} for all $T$. 

Estimating the continuum limit is necessary for making possible a meaningful comparison 
between the expansion coefficients considered in section~\ref{sec:3_1} 
and the generalized quark number susceptibilities $\chi_{j_u,j_d}$. 
In fact, they are closely related~\cite{Shuryak}, e.~g.~the expansion 
coefficients of flavor diagonal and off-diagonal susceptibilities $\chi_{uu}$ 
and $\chi_{ud}$ can be expressed in terms of $\chi_{j_u,j_d}$ via 
\begin{eqnarray}
c_2^{uu} & = & \frac 12 \frac{\chi_{2,0}}{T^2} \,,\\
c_2^{ud} & = & \frac 12 \frac{\chi_{1,1}}{T^2} \,,\\
\label{eq.27}
c_4^{uu} & = & \frac{1}{24} \left(\chi_{4,0}+2\chi_{3,1}+\chi_{2,2}\right) \,,\\
\label{eq.28}
c_4^{ud} & = & \frac{1}{24} \left(2\chi_{3,1}+2\chi_{2,2}\right) \,.
\end{eqnarray}
Within the extended QPM, we find from Eq.~(\ref{equ:coeff1}) and by 
using Eqs.~(\ref{eq.27}) and~(\ref{eq.28}) 
\begin{eqnarray}
\frac{\chi_{2,0}(T)}{T^2} & = & \frac{d}{\pi^2} \int_0^\infty dk \frac{k^2}{T^3} 
\frac{e^{\epsilon_{0}}}{(e^{\epsilon_{0}}+1)^2} \,, \\
\chi_{1,1}(T) & = & 0 \,, \\
\chi_{4,0}(T) & = & \frac{d}{\pi^2} \int_0^\infty dk \frac{k^2}{T^3} 
\frac{e^{\epsilon_{0}}}{(e^{\epsilon_{0}}+1)^4} 
\bigg\{ e^{2\epsilon_{0}} - 4e^{\epsilon_{0}} + 1 \\ \nonumber 
& & - 
\frac{(e^{2\epsilon_{0}}-1)}{\epsilon_{0}} 
\left(\frac{1}{\pi^2} G^2(T) + 
\frac{T^2}{2} \left.\frac{\partial^2 G^2}{\partial\mu_u^2}
\right|_{\mu_{u,d}=0} \right) \bigg\}\,, \\
\chi_{3,1}(T) & = & - \frac{d}{\pi^2} \int_0^\infty dk \frac{k^2}{T^3} 
\frac{e^{\epsilon_{0}}}{(e^{\epsilon_{0}}+1)^4} \frac{(e^{2\epsilon_{0}}-1)}{\epsilon_{0}} 
\frac{T^2}{2} \left(\frac{1}{2}\left.\frac{\partial^2 G^2}{\partial\mu_q^2}
\right|_{\mu_{q,I}=0} - \left.\frac{\partial^2 G^2}{\partial\mu_u^2}
\right|_{\mu_{u,d}=0} \right) , \\
\chi_{2,2}(T) & = & - \frac{d}{\pi^2} \int_0^\infty dk \frac{k^2}{T^3} 
\frac{e^{\epsilon_{0}}}{(e^{\epsilon_{0}}+1)^4} \frac{(e^{2\epsilon_{0}}-1)}{\epsilon_{0}} 
\frac{T^2}{2} \left(\left.\frac{\partial^2 G^2}{\partial\mu_u^2}
\right|_{\mu_{u,d}=0} - \frac13 \left.\frac{\partial^2 G^2}{\partial\mu_q^2}
\right|_{\mu_{q,I}=0} \right) , 
\end{eqnarray}
where $\left.\frac{\partial^2 G^2}{\partial\mu_q^2}
\right|_{\mu_{q,I}=0}$ and $\left.\frac{\partial^2 G^2}{\partial\mu_u^2}
\right|_{\mu_{u,d}=0}$ are given in Eqs.~(\ref{eq.18}) and~(\ref{eq.18_new}). 
As both derivatives of the effective coupling entering these expressions 
are related with each other in the flavor symmetric case, we find $\chi_{3,1}=0$ 
for all temperatures in the QPM, while $\chi_{2,2}$ is non-zero. Furthermore, 
$\chi_{1,1}=0$ as $c_2^{ud}$ vanishes for all temperatures, while 
$c_4^{ud}$ is non-zero as $\chi_{2,2}$ is non-zero. In particular $\chi_{1,1}$, 
or $c_2^{ud}$, vanishes because flavor-mixing effects, which describe 
the dependence of one quark flavor sector on changes in another one, are 
inherent in the quasiparticle model only via the quasiparticle dispersion 
relations resulting in terms which vanish at $\mu_{u,d}=0$. Qualitatively, 
our findings, in particular the observed deviations in the flavor off-diagonal 
susceptibility coefficients, can be understood from perturbative QCD arguments. 
In a perturbative expansion of the thermodynamic potential different partonic 
sectors start to couple only at order $\mathcal{O}(g^3)$ of the 
QCD running coupling $g$. However, these plasmon term contributions $\propto g^3$ 
are not completely reproduced in a similar expansion of the quasiparticle 
model thermodynamic potential~\cite{Romatschke}. Similar findings, pointing 
to the necessity of properly including flavor-mixing effects for affecting 
the flavor off-diagonal susceptibility, were reported in~\cite{PNJL} 
within a Polyakov loop extended Nambu--Jona-Lasinio model approach. 

In Fig.~4, we exhibit the QPM results for $\chi_{2,0}/T^2$ and $\chi_{4,0}$ 
and compare with the continuum extrapolated lattice QCD 
data from~\cite{GavaiCEP,Gavai1} (circles). 
\begin{figure}[t]
\label{fig.4}
\begin{center}
\includegraphics[scale=0.29,angle=0.]{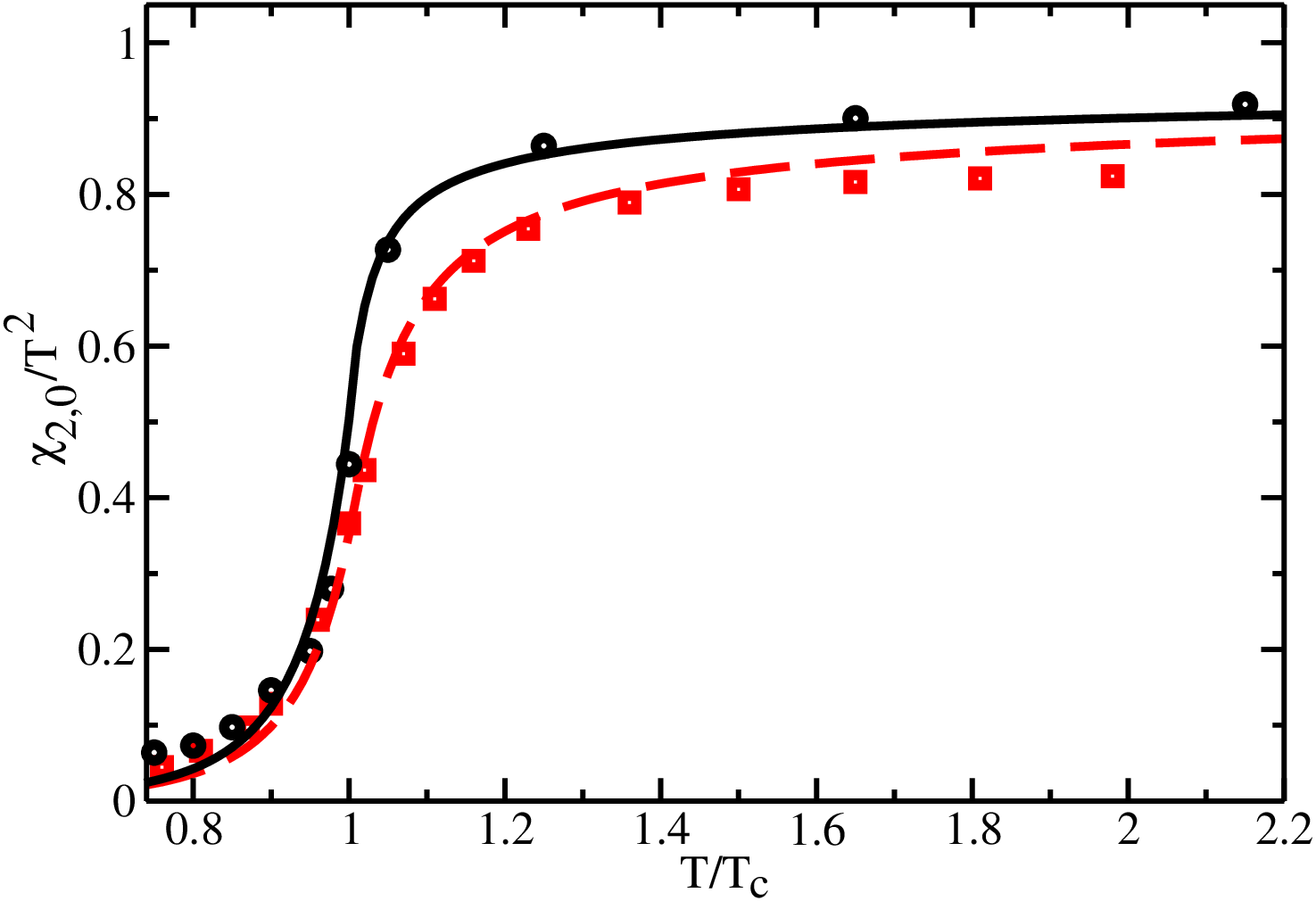}
\hskip 3mm 
\includegraphics[scale=0.29,angle=0.]{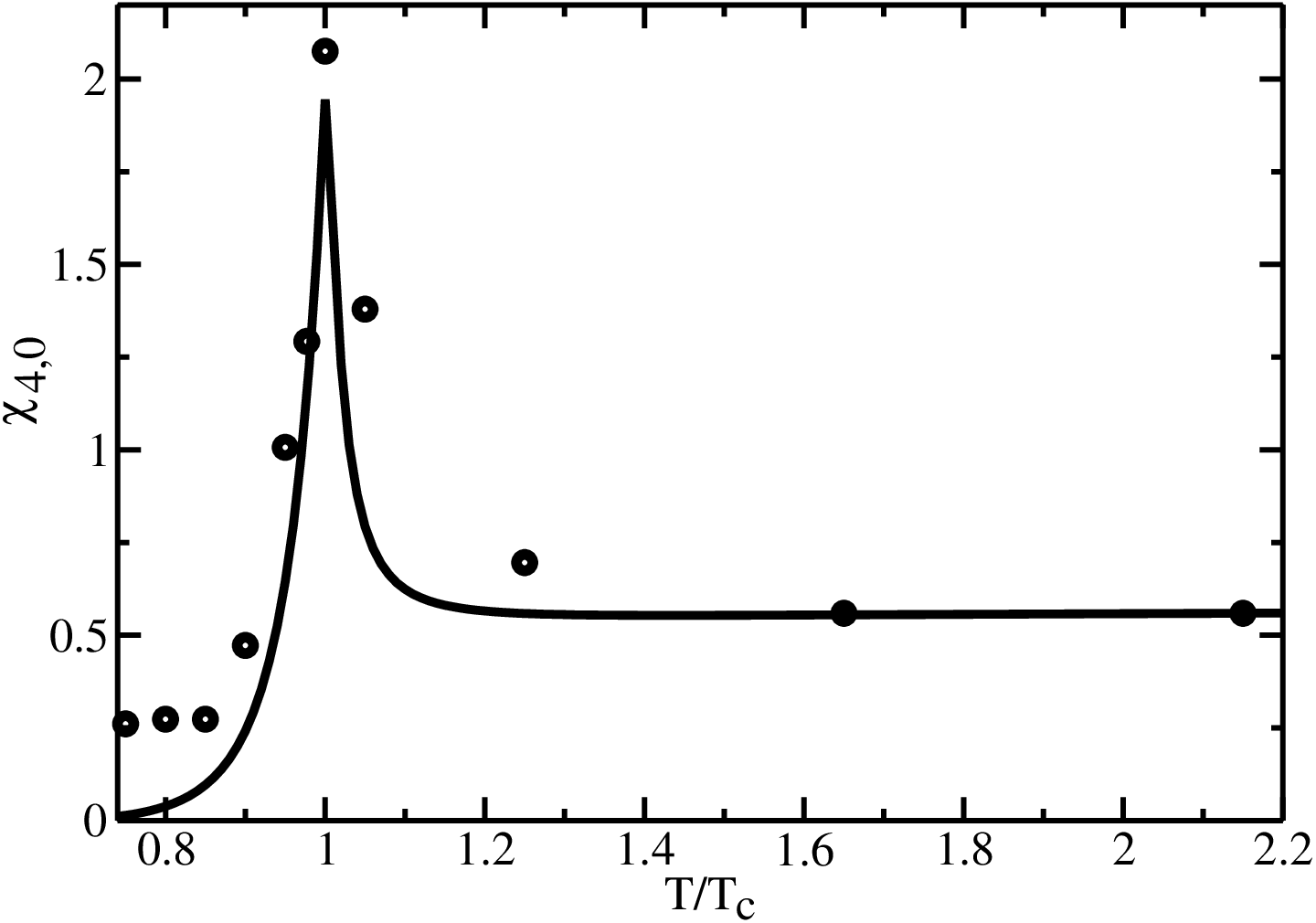}
\caption{(color online). Comparison of QPM results (dashed curves for the parametrization 
employed in section~\ref{sec:3_1} and solid curves for readjusted 
QPM parameters) for the generalized 
quark number susceptibilities $\chi_{2,0}/T^2$ (left panel) 
and $\chi_{4,0}$ (right panel) with the 
continuum extrapolated lattice 
QCD data~\cite{GavaiCEP,Gavai1} (circles) and the lattice QCD data 
for $2c_2^{uu}$ from~\cite{Allton2} (squares).}
\end{center}
\end{figure} 
When using the QPM parameters found in section~\ref{sec:3_1}, the QPM results 
(dashed curve in the left panel of Fig.~4) 
underestimate the lattice QCD data (circles) of $\chi_{2,0}/T^2$. For comparison, we 
also show the lattice QCD data~\cite{Allton2} for $2c_2^{uu}$ (squares), 
where the increasing deviations of the QPM results (dashed curve) 
from the data (squares) 
for increasing temperatures are due to the different quark mass parameters 
used here and in section~\ref{sec:3_1}. (Note that when applying continuum 
limit corrections of about 10\% in the considered temperature range 
to the lattice QCD data~\cite{Allton2} (squares) as 
stated above, both continuum extrapolated lattice QCD data 
sets~\cite{GavaiCEP,Gavai1} (circles) and~\cite{Allton2} would be fairly 
well compatible apart from a narrow interval around $T\approx T_c$ 
such that one unique QPM parametrization would be sufficient.) 
To bridge the data for $\chi_{2,0}$ to $\chi_{4,0}$ by our model, 
we readjust, therefore, the QPM parameters entering $G^2(T)$ in Eq.~(\ref{eff_G}) 
in order to perfectly describe the lattice QCD data~\cite{GavaiCEP,Gavai1} 
(circles) of $\chi_{2,0}/T^2$ 
by using $\lambda=17$, $T_s=0.905\,T_c$ and $b=431$. 
The corresponding QPM results for $\chi_{2,0}/T^2$ and $\chi_{4,0}$ 
are exhibited by solid curves in Fig.~4. Again, very good agreement 
for $T > 0.9\,T_c$ is found. 

%

\subsection{Deconfined $\beta$-stable and electrically neutral quark matter \label{sec:3_3}}

Now, we turn our attention to the discussion of some bulk properties 
of deconfined quark matter of $N_f=2$ dynamical quark flavors 
by means of Taylor series expansions using the generalized quark number 
susceptibilities discussed in the previous section. Clearly, these 
considerations are limited by the range of validity of such an approach, 
say by conservatively guessing the quark flavor chemical potentials 
to be individually restricted by $\mu_{u,d}/T < 1$. Note that we employ 
again $m_{u,d}=0.1\,T_c$ as quark mass parameter. 

Starting from the definition of the excess pressure $\Delta p$ in 
Eq.~(\ref{equ:excess}), including only terms up to $j_u+j_d=4$, the net 
baryon density $n_B$, suppressing the explicit notation of 
the temperature dependence inherent 
in the generalized quark number susceptibilities, reads 
\begin{eqnarray}
 \label{eq.densu}
 n_B(\mu_u,\mu_d) & = & \frac13 \bigg\{ (\chi_{2,0}+\chi_{1,1})(\mu_u+\mu_d) + 
 \left(\frac{\chi_{4,0}}{3!}+\frac{\chi_{3,1}}{3!}\right)(\mu_u^3+\mu_d^3) \\
 \nonumber
 & & + \frac12(\chi_{3,1}+\chi_{2,2})(\mu_u^2\mu_d+\mu_u\mu_d^2) \bigg\} \,,\\
 n_B(\mu_B,\mu_I) & = & \frac19 \bigg\{2(\chi_{2,0}+\chi_{1,1})\mu_B + 
 \frac19\left(\frac13\chi_{4,0}+\frac43\chi_{3,1}+\chi_{2,2}\right)\mu_B^3 \\ 
 \nonumber
 & & + 
 (\chi_{4,0}-\chi_{2,2})\mu_B\mu_I^2 \bigg\} \,.
\end{eqnarray}
Thus, the net baryon density simultaneously depends on two independent chemical 
potentials, $\mu_u$ and $\mu_d$ (or, equivalently, $\mu_B$ and $\mu_I$). This is similarly 
the case for a non-interacting gas of gluons and massless quarks with two 
independent quark flavor chemical potentials. Only in the special case of 
$\mu_I=0$, i.~e.~$\mu_u=\mu_d=\mu_q$, $n_B$ is a function of one chemical 
potential $\mu_B$ alone ensuring constant net baryon density for constant 
baryo-chemical potential. In general, however, a detailed knowledge about 
the dependence on different quark chemical potentials is required, when 
discussing baryon density effects on the EoS. This is illustrated in 
Fig.~5, where the scaled net baryon density is exhibited for 
\begin{figure}[t]
\label{fig.5}
\begin{center}
\includegraphics[scale=0.29,angle=0.]{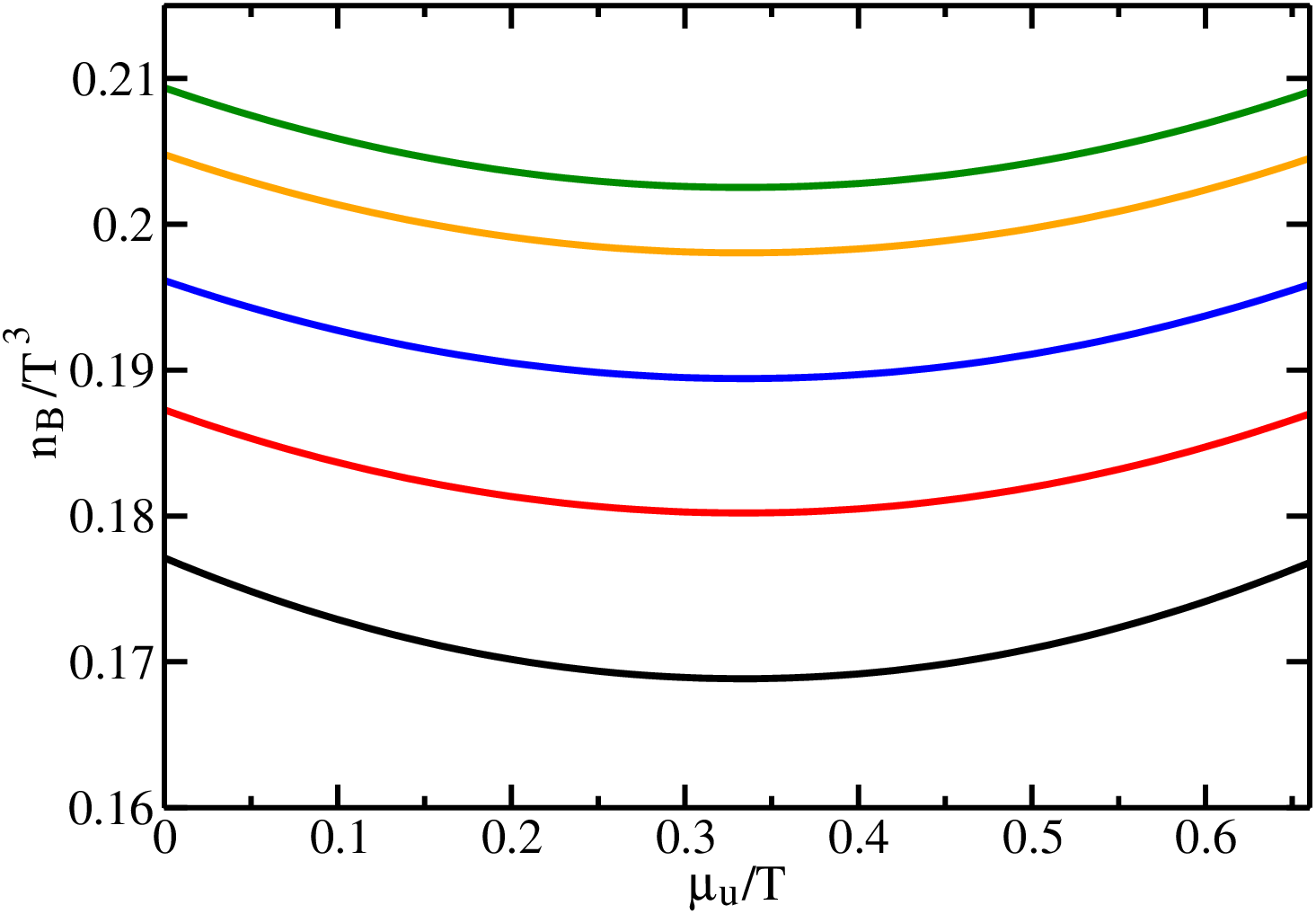}
\caption{(color online). Scaled net baryon density $n_B/T^3$ from Eq.~(\ref{eq.densu}) for 
constant $\mu_B/T=1$ as a function of $\mu_u/T$ for constant 
temperatures $T/T_c=2,\,1.5,\,1.2,\,1.1,\,1.05$ from top 
to bottom.}
\end{center}
\end{figure} 
constant $\mu_B$ and constant temperatures. As by definition 
$\mu_d = \frac23\mu_B-\mu_u$, one chemical potential in 
Eq.~(\ref{eq.densu}) can be replaced. We chose $\mu_B/T = 1$ such that 
$\mu_u/T+\mu_d/T = \frac23$, ensuring that these considerations stay 
within the range of validity of the employed Taylor expansion 
approach. The minimum at $\mu_u/T=\mu_d/T=\frac13$ 
exhibits the value of $n_B/T^3$ for one independent quark chemical 
potential. $n_B$ for $\mu_B=T$ drops by $3.3$\% at $T=2\,T_c$ and by 
$4.6$\% at $T=1.05\,T_c$ when changing $\mu_u/T$ from $0$ to 
$\frac13$. Accordingly, one is tempted to consider the detailed 
knowledge about the individual $\mu_u$ and $\mu_d$ dependencies as not so important. 

However, there are physical situations, where the corresponding side 
conditions require the separate knowledge about the non-trivial $\mu_u$ and 
$\mu_d$ dependencies of bulk thermodynamic quantities. 
First, we consider curves of constant 
$\mu_B$, which are given by the linear relation $\mu_d = \frac23\mu_B-\mu_u$ 
(see short-dashed 
curve in the left panel of Fig.~6 with $\mu_B/T = 1$). The individual 
net quark number densities read $n_u = \chi_{2,0}\mu_u + \chi_{1,1}\mu_d + \frac{\chi_{4,0}}{3!}\mu_u^3 + 
 \frac{\chi_{3,1}}{2}\mu_u^2\mu_d + \frac{\chi_{3,1}}{3!}\mu_d^3 + 
 \frac{\chi_{2,2}}{2}\mu_u\mu_d^2$ and $n_d = \chi_{2,0}\mu_d + \chi_{1,1}\mu_u + \frac{\chi_{4,0}}{3!}\mu_d^3 + 
 \frac{\chi_{3,1}}{2}\mu_u\mu_d^2 + \frac{\chi_{3,1}}{3!}\mu_u^3 + 
 \frac{\chi_{2,2}}{2}\mu_u^2\mu_d$. Since in the QPM $\chi_{1,1}=\chi_{3,1}=0$, 
lines of constant $n_u$ or $n_d$ are approximately given by lines of 
constant $\mu_u$ or $\mu_d$, i.~e.~simply vertical or horizontal lines 
in the left panel of Fig.~6 (not displayed). 
(Only at temperatures $T\approx T_c$, where 
$\chi_{2,2}$ is non-negligible, the simple pattern is deformed 
somewhat.) This situation is completely different when considering 
constant scaled net baryon densities as depicted by the solid curve in Fig.~6 
(left panel) for $n_B/T^3 = 0.187$ at $T = 1.1\,T_c$ unravelling the non-trivial 
dependence of $\mu_d$ on $\mu_u$ in contrast to constant $\mu_B$. 
\begin{figure}[t]
\label{fig.6}
\begin{center}
\includegraphics[scale=0.29,angle=0.]{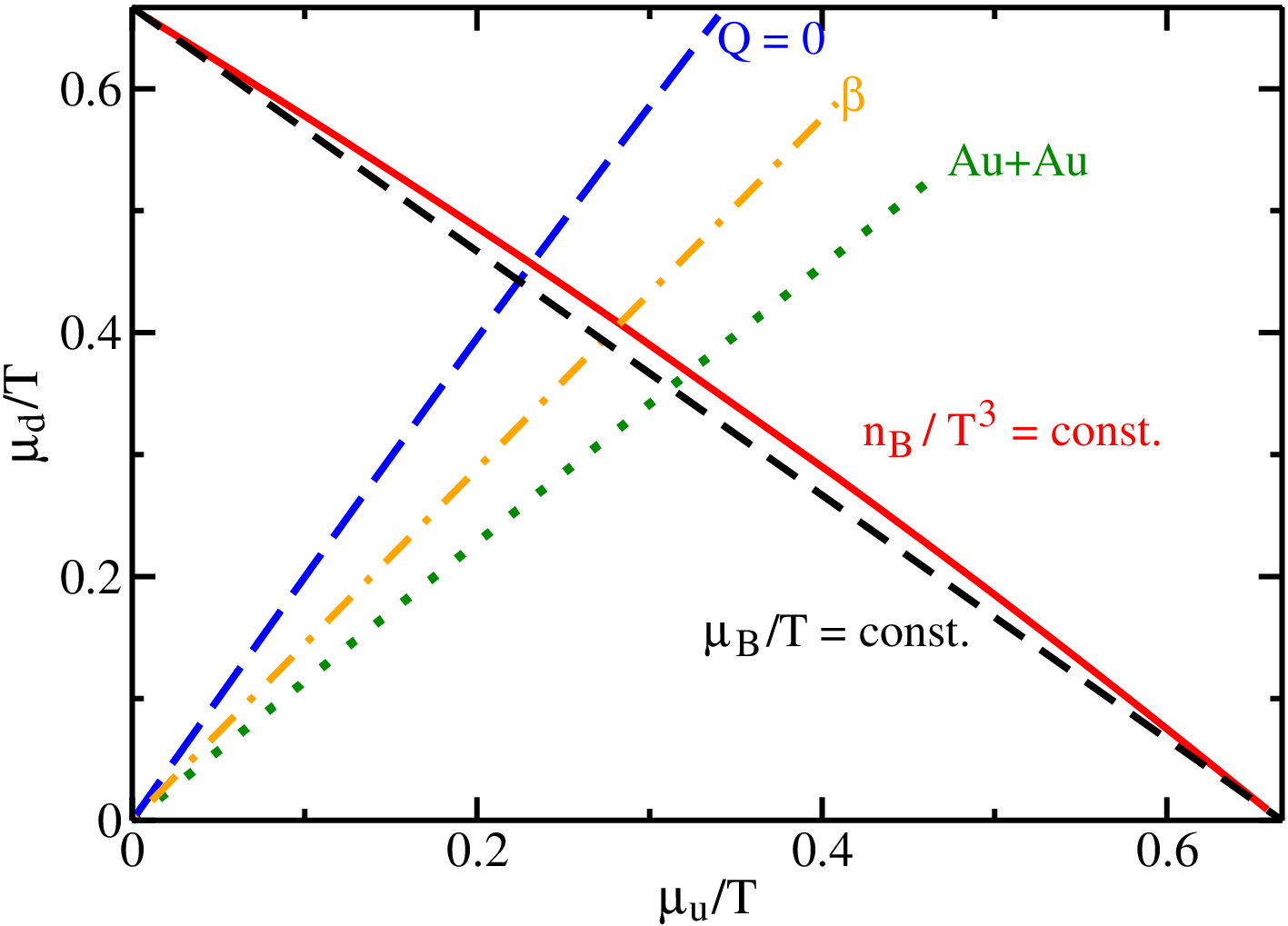}
\hskip 3mm 
\includegraphics[scale=0.29,angle=0.]{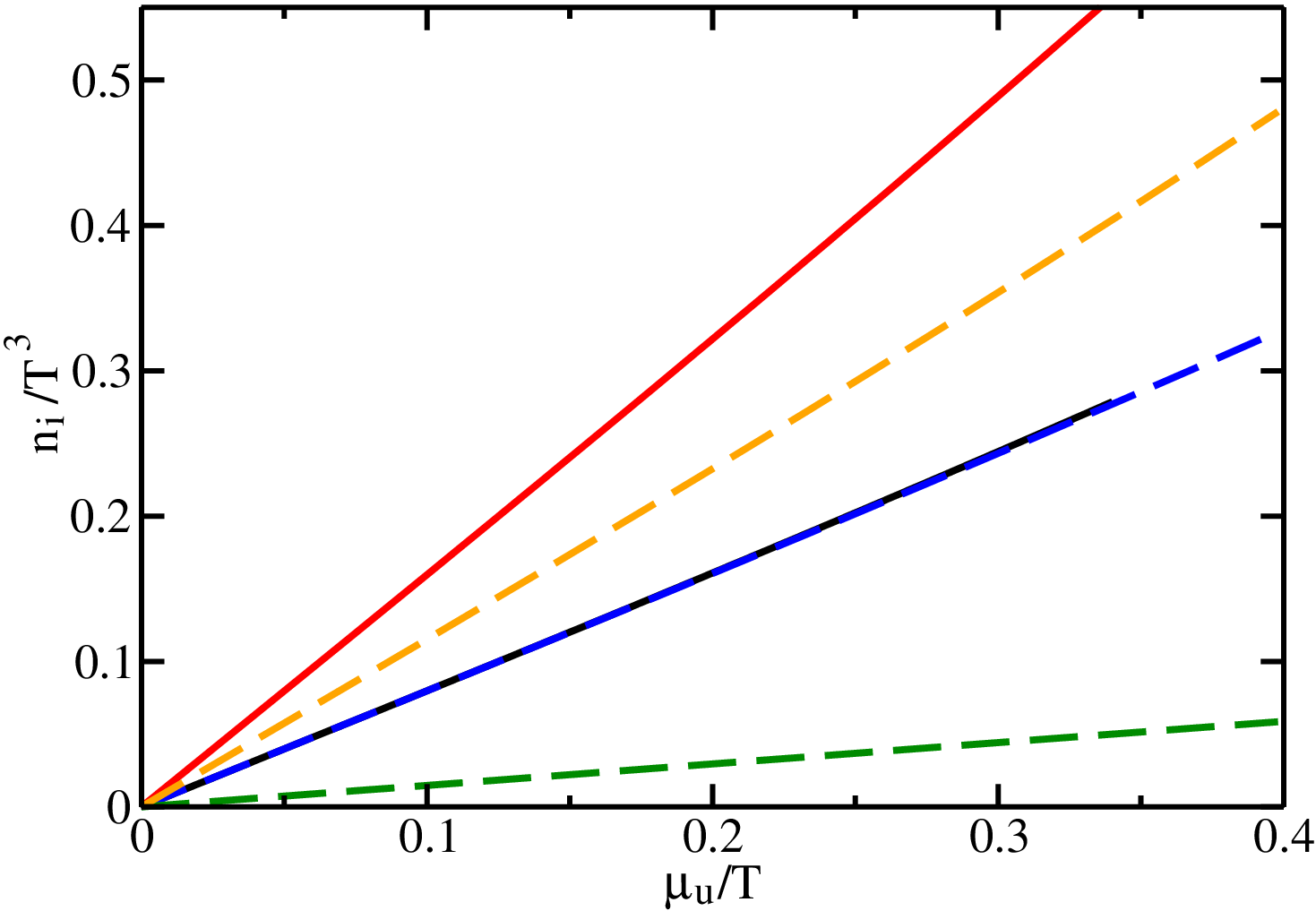}
\caption{(color online). Left: Dependence $\mu_d(\mu_u)$ for various 
side conditions or physical situations. 
$\mu_B=T$ is depicted by the short-dashed curve, whereas constant $n_B/T^3=0.187$ 
holds along the solid curve where $\mu_B\ge T$. 
Electric charge neutrality is given along the long-dashed curve for pure $N_f=2$ 
quark matter, while the dash-dotted curve includes additionally electrons, 
imposing $\beta$-equilibrium. The dotted curve reflects the situation in 
$Au+Au$ heavy-ion collisions. (The curves end where $\mu_u/T+\mu_d/T \ge 1$.) 
Right: Scaled net number densities as functions 
of $\mu_u/T$ demanding electric charge neutrality either for pure $N_f=2$ quark matter (solid 
curves, $n_d/T^3$ - top, $n_u/T^3$ - bottom) or for including electrons and 
requiring $\beta$-equilibrium (dashed curves, $n_d/T^3$, $n_u/T^3$, $n_e/T^3$ from top to bottom). 
For $T=1.1\,T_c$.}
\end{center}
\end{figure} 
In fact, here $\mu_B/T > 1$ except for the case when $\mu_u/T=0$ or $\mu_d/T=0$. 

In heavy-ion collisions one often relates the quantum numbers of the entrance 
channel with the ones of the emerging fireball. Isospin-symmetric nuclear matter, 
for instance, is characterized by an electric charge per baryon ratio of 1:2. This 
translates into $\frac23 n_u - \frac13 n_d = \frac12 n_B$ which is fulfilled for 
$\mu_d=\mu_u$, i.~e.~simply a diagonal line in the left panel of Fig.~6 (not displayed). 
Discussing, instead, gold on gold collisions, the electric charge per baryon ratio is 
approximately $0.4$. The corresponding dependence $\mu_d(\mu_u)$ for $T=1.1\,T_c$ is 
depicted by the dotted curve in the left panel of Fig.~6. 
Another important issue concerns electric charge neutrality in bulk matter. 
In pure $N_f=2$ quark matter, electric charge neutrality would require 
$\frac23 n_u - \frac13 n_d = 0$. The according 
dependence $\mu_d(\mu_u)$ is depicted in Fig.~6 (left panel) by the long-dashed 
curve, again for $T=1.1\,T_c$. More relevant for hypothetical very hot 
neutron star matter in a deconfined state is $\beta$-equilibrium. Flavor changing 
weak currents give rise to the balance equation 
$d\leftrightarrow u + e + \bar{\nu_e}$, i.~e.~in weak interaction 
equilibrium $\mu_e=\mu_d-\mu_u$, as the produced neutrinos are supposed 
to leave the star and do not participate in the balance. The electron 
net density is approximated by $n_e=\frac13 \mu_eT^2 + 
\frac{1}{3\pi^2}\mu_e^3$, and electrically 
neutral bulk matter is determined by $\frac23 n_u - \frac13 n_d - n_e= 0$. 
The corresponding dependence 
$\mu_d(\mu_u)$ is depicted by the dash-dotted curve in Fig.~6 (left panel) for 
$T=1.1\,T_c$. The $d$ quark net number density 
decreases by requiring $\beta$-equilibrium, 
demanding also a non-zero electron density for electrically neutral 
bulk matter (see Fig.~6 right panel); $n_u$ is not affected when 
including electrons and $\beta$-equilibrium. This is in contrast to findings for 
the cold color-flavor locked phase of QCD~\cite{Wilczek} for $N_f=2+1$ dynamical 
quarks, where no electrons are required. 

The discussion can easily be extended to the physically relevant case of two light 
(up and down) and one heavier (strange) quarks, considering again two independent 
quark chemical potentials for the light quarks, $\mu_l=\mu_u=\mu_d$, and for 
the strange quark, $\mu_s$. Recently, first-principle lattice QCD data for this 
case became available~\cite{Karschcutoff,Miao}. A detailed comparison 
of the properly extended quasiparticle model with these lattice results and, in 
particular, a discussion of finite baryon density effects on the EoS 
relevant for the hydrodynamical description of the expansion stage of 
heavy-ion collisions demands further studies. 

\section{Summary and discussion \label{sec:conclusions} } 

The focus of the present paper is an analysis of isovector and various flavor (off-)diagonal
susceptibilities for two-flavor QCD by comparing the extended quasiparticle 
model with lattice QCD data~\cite{GavaiCEP,Allton2,Gavai1}. The model includes
the same quark mass parameters $m_i$ as used in these lattice simulations.
(Basically, one could accomplish also a chiral extrapolation.
However, the effective coupling $G^2(T)$ may implicitly depend on these masses. This deserves
separate investigations.) A crucial point to be kept in mind concerns
finite size effects. The lattice QCD calculations \cite{GavaiCEP,Allton2,Gavai1} are
performed on grids with finite temporal and spatial extension, 
while our phenomenological model is formulated in
the thermodynamic continuum limit. The use of an improved action in
\cite{Allton2} lets us hope that the finite size effects are
sufficiently small to make a direct comparison meaningful. 
In contrast, the lattice QCD data of~\cite{GavaiCEP,Gavai1} require 
severe continuum extrapolation factors. In so far, the 
comparison of our extended QPM with these data is less direct. 

Having these limitations in mind, we emphasize the good agreement of
our model with the lattice QCD data for $c_{2,4}$, $c_{2,4}^I$, $c_{2,4}^{uu}$
and $c_4^{ud}$ as well as for the related generalized quark number susceptibilities. 
We consider this successful
comparison as encouraging. A conclusion is that quasiparticle
excitations, with a mass gap also in the chiral limit, are able to
explain those features of the strongly coupled quark-gluon medium
which are encoded in the mentioned coefficients. In particular,
baryon density effects are probed by these coefficients. 
The baryon charge is carried by quasi-quark excitations, in contrast 
to models~\cite{Shuryak}, where di-quark and three-quark modes 
carry a substantial fraction of the baryon charge. 
Furthermore, in several physical situations, like relativistic heavy-ion collisions 
or in hot proto-(quark) neutron stars, the various mentioned coefficients 
are needed to implement the adequate side conditions. 

We have applied our model also for $T < T_c$. Formally, the
description of the lattice QCD data below $T_c$ requires fairly large
values of the effective coupling $G^2(T)$. (An alternative description
could rely on strongly increasing correlations which are beyond the
presently employed 
approach~\cite{Bluhm_EPJC}.) The
corresponding excitations become very massive, ranging to hadronic
mass scales. It turns out that a few massive excitations
reproduce fairly well some of the lattice QCD data 
within the interval $0.8\, T_c - T_c$. This is
numerically not too distinct from the hadron resonance gas model,
where one may regroup several resonances into a few representative
effective excitations. (Vice versa, we mention that the resonance gas
model \cite{Allton1,Allton2,Redlich} coincides with lattice QCD data also
slightly above $T_c$; for an even more extreme point of view we
refer the interested reader to \cite{Blaschke}.) In this respect, it
is conceivable that several models with fairly distinct assumptions
may equally well reproduce the same lattice QCD data on thermodynamic bulk
properties - examples 
are~\cite{Shuryak,Sasaki1,Sasaki2,Ghosh,Roessner,Bannur,Bluhm_PoS,Weise,Toneev,Biro}.
Only correlators and spectral properties of
the excitations can unreveal their real nature in the strongly
interacting system. 

On the other hand, the coefficient $c_2^{ud}$, 
and accordingly $\chi_{1,1}$, is poorly described. This
may be a hint for missing modes or degrees of freedom in our model. 
Qualitatively, our findings can be understood since flavor-mixing effects, 
which are important for the correct description of the flavor off-diagonal 
susceptibility, are not explicitly inherent in our quasiparticle model, 
but only implicitly via the quasiparticle dispersion relations. 
Progressing lattice QCD calculations are welcome to resolve this
issue and to get more confidence in the baryon number carrying
modes (cf.~discussions in~\cite{Gavaisigns1,Karsch}). 
Also, the slight deviations between our model and the data very
close to $T_c$ may signal a deficit of our quasiparticle picture.
Nevertheless, considering our phenomenological model as useful parametrization
of lattice QCD results, it may serve as QCD-based input for hydrodynamical
calculations for the expansion dynamics of matter created in ultra-relativistic
heavy-ion collisions, cf. \cite{Bluhm_PRC}. 

Finally, we stress that the utilized Taylor expansion technique is sensitive
to the region $\mu_{u,d} \rightarrow 0$. QCD critical point effects
at larger values of $\mu_{u,d}$ may not be catched in such an
approach. For a phenomenological procedure to supplement our model
by critical point features see \cite{Bluhm_QM_POS06}.

In summary, we extend our quasiparticle model towards two independent 
chemical potentials. This allows for the determination of various 
susceptibilities. We find an impressive agreement (with the exception 
of two numerically small flavor off-diagonal susceptibility coefficients) 
with lattice QCD data. Since a special set of susceptibilities also 
provides the Taylor expansion coefficients of the baryon-driven 
excess pressure, we argue that our phenomenological quasiparticle 
model catches relevant modes for the equation of state at 
non-zero net baryon density. It may be used, therefore, 
for the future determination of higher-order Taylor expansion coefficients 
which become increasingly important at larger net baryon densities. 

\subsection*{Acknowledgements}

We gratefully acknowledge discussions with E. Laermann, F. Karsch,
R.~V.~Gavai and S.~Gupta. The work is supported by BMBF 06DR136, GSI-FE, EU I3HP.

\section*{Appendix A}

The pressure $p(T,\mu_u,\mu_d)$ as primary thermodynamic potential 
of our model is constructed by assuming a quasiparticle picture 
via 
\begin{equation}
\label{equ:tdPot}
p(T,\mu_u,\mu_d) = \sum_{l=u,d,g} p_l(T,\mu_u,\mu_d) - B(\Pi_{u,d,g}[T, \mu_u, \mu_d]), 
\end{equation}
where $B$ is to be determined as line
integral from thermodynamic consistency conditions and the
stationarity condition $\delta p /\delta \Pi_j=0$ resulting in
$\partial p_j / \partial \Pi_j = \partial B / \partial \Pi_j$.
The partial pressures $p_l$ of included excitations $l$ referring to 
$u$ quarks, $d$ quarks and gluons ($g$) read 
\begin{equation}
\label{equ:QPM2}
p_l = \epsilon_l d_l T \int \frac{d^3 k}{(2\pi)^3} 
\left[\ln \left(1+\epsilon_l e^{-(\omega_l-\mu_l)/T}\right) + 
\ln \left(1+\epsilon_l e^{-(\omega_l+\mu_l)/T}\right)\right] ,
\end{equation}
where the dispersion relations $\omega_l=\omega_l(T,\mu_u,\mu_d)$ 
are given in Eqs.~(\ref{eq.5}) and~(\ref{eq.7}), 
$\epsilon_l$ is $+1$ ($-1$) for fermions (bosons), $d_l$ refers 
to the spin (polarization) and color degeneracies of quasiquarks and quasigluons reading 
$d_u=d_d=2\,N_c$ and $d_g=N_c^2-1$, and $\mu_g=0$. 
In this way, we count left-handed transversal quasigluons 
as anti-particles of the right-handed ones. 
These structures emerge from the underlying two-loop QCD 
{\boldmath$\Phi$} functional~\cite{Baym,Blaizot} by imposing 
formal manipulations such as neglecting finite width 
effects in the considered asymptotic HTL approximations of the one-loop 
self-energies, and neglecting (anti)plasmino and longitudinal gluon excitations 
as well as Landau damping~\cite{Bluhm_EPJC}. 
While $p$ is highly non-perturbative with respect to the effective coupling
$G^2$ entering the self-energy expressions, it is this 
phenomenologically introduced coupling which enables the model 
to go beyond the {\boldmath$\Phi$}-derivable approximations in \cite{Blaizot}. 

The entropy density expression entering the generalized Peshier equations 
in Eqs.~(\ref{equ:QPM31}-\ref{equ:QPM33}) reads 
$s = \sum_{l=u,d,g} s_l = \partial p / \partial T$ with 
\begin{eqnarray}
\label{equ:QPM7}
s_i & = & d_i \int \frac{d^3 k}{(2\pi)^3} \left[\ln \left(1 + e^{-(\omega_i-\mu_i)/T}\right) 
+ 
\frac{(\omega_i - \mu_i)/T}{\left(e^{(\omega_i-\mu_i)/T}+1\right)} + (\mu_i\rightarrow -\mu_i)\right], \\
\label{equ:QPM8}
s_g & = & -2 d_g \int \frac{d^3 k}{(2\pi)^3} \left[\ln \left(1 - e^{-\omega_g/T}\right) - 
\frac{\omega_g/T}{\left(e^{\omega_g/T}-1\right)}\right],
\end{eqnarray}
where $i=u,d$ and $\mu_u=\mu_q+\mu_I$, $\mu_d=\mu_q-\mu_I$. 
This additivity in the contributions $s_l$ of the various parton
species is anchored in the underlying two-loop QCD 
{\boldmath$\Phi$} functional~\cite{Bluhm_EPJC,Blaizot,Vanderheyden}. 

\section*{Appendix B}

The coefficients entering Eqs.~(\ref{eq.11}~-~\ref{eq.13}) read
\begin{eqnarray} 
A_1 & = & \mathcal{I}_3 \frac13 \left[ 2 T^2 +
\frac{3}{2\pi^2} \left( \mu_q^2+\mu_I^2\right)\right] + 
\mathcal{I}_4\frac13 \left[ T^2+
\frac{1}{\pi^2}(\mu_q+\mu_I)^2\right]  \\ \nonumber
& & + \mathcal{I}_5 \frac13 \left[ T^2+\frac{1}{\pi^2} (\mu_q-\mu_I)^2\right] \,,\\
B_1 & = & - \mathcal{I}_1 \frac13 \left[ T^2+\frac{1}{\pi^2}(\mu_q+\mu_I)^2\right] + 
\mathcal{I}_2 \frac13 \left[ T^2+\frac{1}{\pi^2}(\mu_q-\mu_I)^2\right] \,,\\
C_1 & = & - \mathcal{I}_3 \frac{1}{\pi^2} G^2\mu_I - \mathcal{I}_4 
\frac13 \frac{2}{\pi^2}[\mu_q+\mu_I] G^2 
+ \mathcal{I}_5 \frac13 \frac{2}{\pi^2}[\mu_q-\mu_I] G^2 \\ \nonumber 
& & + \mathcal{I}_1 \left( 2 \xi_u^2T + \frac23 T G^2 \right) - 
\mathcal{I}_2 \left( 2 \xi_d^2 T + \frac23 T G^2\right) \,,\\
A_2 & = & \mathcal{I}_3 \frac13 \left[ 2 T^2 +
\frac{3}{2\pi^2}\left(\mu_q^2+\mu_I^2\right)\right] + 
\mathcal{I}_4 \frac13 \left[ T^2+
\frac{1}{\pi^2}(\mu_q+\mu_I)^2\right] \\ \nonumber
& & + \mathcal{I}_5 \frac13 \left[T^2+\frac{1}{\pi^2}(\mu_q-\mu_I)^2\right] \,,\\
B_2 & = & - \mathcal{I}_1 \frac13 \left[T^2+\frac{1}{\pi^2}(\mu_q+\mu_I)^2 \right] - 
\mathcal{I}_2 \frac13 \left[T^2+\frac{1}{\pi^2}(\mu_q-\mu_I)^2 \right] \,,\\
C_2 & = & - \mathcal{I}_3 \frac{1}{\pi^2} G^2 \mu_q - 
\mathcal{I}_4 \frac13 \frac{2}{\pi^2}[\mu_q+\mu_I] G^2 
- \mathcal{I}_5 \frac13 \frac{2}{\pi^2}[\mu_q-\mu_I] G^2 \\ \nonumber 
& & + \mathcal{I}_1 \left( 2 \xi_u^2 T + \frac23 T G^2 \right) + 
\mathcal{I}_2 \left( 2 \xi_d^2 T + \frac23 T G^2 \right) \,,\\
A_3 & = & \mathcal{I}_1 \frac13 \left[T^2+\frac{1}{\pi^2}(\mu_q+\mu_I)^2\right] \,,\\
B_3 & = & \mathcal{I}_2 \frac13 \left[T^2+\frac{1}{\pi^2}(\mu_q-\mu_I)^2\right] , 
\end{eqnarray}
where the phase-space integrals $\mathcal{I}_{k}$ are given by 
\begin{eqnarray} 
\mathcal{I}_1 & = & \frac{\partial n_u}{\partial\Pi_u} = \frac{d_u}{2\pi^2}\int_0^\infty 
dk \frac{k^2}{2\omega_u T} \left( 
\frac{e^{(\omega_u+\mu_q+\mu_I)/T}}{(e^{(\omega_u+\mu_q+\mu_I)/T}+1)^2} - 
\frac{e^{(\omega_u-\mu_q-\mu_I)/T}}{(e^{(\omega_u-\mu_q-\mu_I)/T}+1)^2}\right) \,, \\
\mathcal{I}_2 & = & \frac{\partial n_d}{\partial\Pi_d} = \frac{d_d}{2\pi^2}\int_0^\infty 
dk \frac{k^2}{2\omega_d T} \left( 
\frac{e^{(\omega_d+\mu_q-\mu_I)/T}}{(e^{(\omega_d+\mu_q-\mu_I)/T}+1)^2} - 
\frac{e^{(\omega_d-\mu_q+\mu_I)/T}}{(e^{(\omega_d-\mu_q+\mu_I)/T}+1)^2}\right) \,, \\
\label{equ:add1}
\mathcal{I}_3 & = & \frac{\partial s_g}{\partial\Pi_g} = -\frac{d_g}{\pi^2}\int_0^\infty 
dk \frac{k^2}{2T^2} 
\frac{e^{\omega_g/T}}{(e^{\omega_g/T}-1)^2} \,, \\
\label{equ:add2}
\mathcal{I}_4 & = & \frac{\partial s_u}{\partial\Pi_u} = -\frac{d_u}{2\pi^2}\int_0^\infty 
dk \frac{k^2}{2\omega_u T^2} \left( 
\frac{(\omega_u+\mu_q+\mu_I)e^{(\omega_u+\mu_q+\mu_I)/T}}{(e^{(\omega_u+\mu_q+\mu_I)/T}+1)^2} + 
(\mu_{q,I}\rightarrow -\mu_{q,I})\right), \\
\label{equ:add3}
\mathcal{I}_5 & = & \frac{\partial s_d}{\partial\Pi_d} = -\frac{d_d}{2\pi^2}\int_0^\infty 
dk \frac{k^2}{2\omega_d T^2} \left( 
\frac{(\omega_d+\mu_q-\mu_I)e^{(\omega_d+\mu_q-\mu_I)/T}}{(e^{(\omega_d+\mu_q-\mu_I)/T}+1)^2} + 
(\mu_{q,I}\rightarrow -\mu_{q,I})\right). 
\end{eqnarray}
In Eqs.~(\ref{equ:A10}) and~(\ref{equ:A11}), $\mu_q$ and $\mu_I$ in the 
phase-space integrals $\mathcal{I}_{k}$ are replaced by 
$\mu_u=\mu_q+\mu_I$ and $\mu_d=\mu_q-\mu_I$. 

\section*{Appendix C}

The coefficients in Eqs.~(\ref{eq.14}) and~(\ref{eq.15}) read 
\begin{eqnarray} 
\label{equ:A6}
\mathcal{A}_1 = \mathcal{A}_2 & = & \frac13 \mathcal{I}_3
\left[ 2 T^2+
\frac{3}{4\pi^2}\left(\mu_u^2+\mu_d^2\right)\right] + 
\frac13 \mathcal{I}_4  \left[ T^2 + \frac{\mu_u^2}{\pi^2}\right] + 
\frac13 \mathcal{I}_5 \left[ T^2+\frac{\mu_d^2}{\pi^2}\right] \,, \\
\label{equ:A7}
\mathcal{B}_1 = \mathcal{B}_2 & = & - \frac13 \mathcal{I}_1 
\left[T^2+\frac{\mu_u^2}{\pi^2}\right] 
\end{eqnarray}
and 
\begin{eqnarray} 
\label{equ:A8}
\mathcal{C}_1 & = & \mathcal{I}_1 \left( 2 \xi_u^2 T + \frac23 T G^2\right) 
- \mathcal{I}_3 \frac{1}{2\pi^2}G^2\mu_u - \mathcal{I}_4 
\frac{2}{3 \pi^2}\mu_u G^2 \,, \\
\label{equ:A9}
\mathcal{C}_2 & = & \mathcal{I}_1 \frac{(T^2+\mu_u^2/\pi^2)}{(T^2+\mu_d^2/\pi^2)} 
\left(2 \xi_d^2 T + \frac23 T G^2 \right) - \mathcal{I}_3 \frac{1}{2\pi^2} G^2 \mu_d 
\frac{\mathcal{I}_1}{\mathcal{I}_2} \frac{(T^2+\mu_u^2/\pi^2)}{(T^2+\mu_d^2/\pi^2)} \\ 
\nonumber 
& & - \mathcal{I}_5 \frac{2}{3 \pi^2}\mu_d G^2 
\frac{\mathcal{I}_1}{\mathcal{I}_2} 
\frac{(T^2+\mu_u^2/\pi^2)}{(T^2+\mu_d^2/\pi^2)}\,, 
\end{eqnarray}
where $\mu_q$ and $\mu_I$ in the phase-space integrals 
$\mathcal{I}_{k}$ defined in Appendix B 
have to be substituted by 
$\mu_u=\mu_q+\mu_I$ and $\mu_d=\mu_q-\mu_I$. 
The coefficients in Eq.~(\ref{eq.18}) read
\begin{eqnarray} 
\label{equ:QPM50}
\mathcal{N} & = & T^2 \left( \frac23 \mathcal{I}_3  + 
\frac13  \mathcal{I}_4  + \frac13 \mathcal{I}_5 \right) , \\
\mathcal{N}_1 & = & \frac{d_u}{2\pi^2} \int_0^\infty dk \frac{k^2}{T^2\omega_u} 
\left(\frac{e^{\omega_u/T}}{(e^{\omega_u/T}+1)^2} - 
2 \frac{e^{2\omega_u/T}}{(e^{\omega_u/T}+1)^3}\right) ,\\
\mathcal{N}_2 & = & \frac{d_d}{2\pi^2} \int_0^\infty dk \frac{k^2}{T^2\omega_d} 
\left(\frac{e^{\omega_d/T}}{(e^{\omega_d/T}+1)^2} - 
2 \frac{e^{2\omega_d/T}}{(e^{\omega_d/T}+1)^3}\right) \,,
\end{eqnarray}
where $\mathcal{I}_{3,4,5}$ as well as $\omega_{u,d}$ have to 
be taken at $\mu_q=\mu_I=0$ or, equivalently, $\mu_u=\mu_d=0$. 

\section*{Appendix D}

Let us first briefly discuss an implication of the requirement 
$\mu_{u,d}\ll\pi T$ needed for the consistency of 
Eqs.~(\ref{eq.14}) and~(\ref{eq.15}). Second-order susceptibility 
coefficients depend on $G^2$ evaluated at $\mu_{u,d}=0$, 
while fourth-order coefficients depend on $G^2$ and 
$\partial^2G^2/\partial\mu_{u,d}^2$ at $\mu_{u,d}=0$. 
In general, $n$-th order derivatives of 
$G^2$ require up to and including ($n-1$)-st derivatives of 
${\cal C}_1$ or ${\cal C}_2$. This implies that up to and including 
third-order the derivatives of the 
effective coupling can trustfully be taken. Therefore, second- and 
fourth-order susceptibility coefficients and related quantities 
are uniquely determined. However, 
$\left.\frac{\partial^4 G^2}{\partial\mu_u^4}\right|_{\mu_u=\mu_d=0}$ 
and higher orders cannot be evaluated uniquely. 
These derivatives enter, for instance, sixth- and higher-order 
susceptibility coefficients. 

The origin of this insanity 
is the special ansatz for the self-energy parts in the quasi-particle 
dispersion relations in Eqs.~(\ref{eq.5}) and~(\ref{eq.7}), while 
our primary thermodynamic potential in Eq.~(\ref{equ:tdPot}) together 
with~(\ref{equ:QPM2}) should allow for consistency in all orders of 
powers of $\mu_{u,d}$. The reasoning for our ansatz in 
Eqs.~(\ref{eq.5}) and~(\ref{eq.7}) is the contact to one-loop expressions 
for the self-energies~\cite{leBellac}. It has been shown, however, 
in~\cite{Bluhm_immu}, for one (imaginary) chemical potential, that one 
can discard the explicit $\mu^2$ terms in the self-energies and obtains 
an equally suitable description of the lattice QCD results. In other 
words, the stationarity property of the thermodynamic potential $p$, 
involved in our quasiparticle model, causes a robustness against such 
modifications of the employed self-energy parametrizations. 

It happens that for the modified self-energies, 
$\Pi_i=\frac13 T^2G^2(T,\mu_u,\mu_d)$ and 
$\Pi_g=\frac23 T^2G^2(T,\mu_u,\mu_d)$, the coefficients in 
Eqs.~(\ref{eq.14}) and~(\ref{eq.15}) render to 
\begin{eqnarray}
    {\cal A}_1 & = & {\cal A}_2 = \frac{T^2}{3}\left(2{\cal I}_3
    +{\cal I}_4+{\cal I}_5
    \right) \,, \\
    {\cal B}_1 & = & {\cal B}_2 = -\frac{T^2}{3}{\cal I}_1 \,,\\
    {\cal C}_1 & = & {\cal C}_2 = \frac23 TG^2{\cal I}_1 \,,
\end{eqnarray}
(for simplicity, we consider here the chiral limit or, as in 
section~\ref{sec:3_2}, temperature independent bare quark masses). 
I.~e.~the generalized system of flow equations in 
Eqs.~(\ref{eq.11}-\ref{eq.13}) is uniquely solvable without restrictions, 
and $G^2$ and all its derivatives are trustfully obtained, implying 
also a consistent determination of the susceptibility 
coefficients of arbitrary order opening the avenue for future 
investigations. We have checked numerically that the result exhibited for the 
fourth-order coefficient in Fig.~4 is changed by less than 
9\% when changing the self-energy expressions 
(generically a slight down shift of the 
curves occurs). The result for the second-order coefficient 
exhibited in Fig.~4 remains unchanged as it depends only on $G^2$ at 
$\mu_{u,d}=0$ which is not affected by the modification of the 
self-energy parametrizations. Similar statements are applicable for other 
related susceptibilities. Consequently, the results exhibited in 
Figs.~5 and~6 remain effectively unaltered.


\begin{thebibliography}{100}

\bibitem{RHIC_announcement} E.~V.~Shuryak, Prog. Part. Nucl. Phys. {\bf 53}, 273 (2004); 
 Nucl. Phys. A {\bf 750}, 64 (2005). 
\bibitem{McLerran} M.~Gyulassy, and L.~McLerran, Nucl. Phys. A {\bf 750}, 30 (2005). 
\bibitem{Teaney} D.~A.~Teaney, Phys. Rev. C {\bf 68}, 034913 (2003); 
 J. Phys. G {\bf 30}, S1247 (2004); 
 Nucl. Phys. A {\bf 785}, 44 (2007). 
\bibitem{CEP} Proceedings of Critical Point and Onset of Deconfinement - 3rd 
 International Workshop, July 3-6, 2006, Florence, Italy, (Ed.) F.~Becattini; 
 4th International Workshop, July 9-13, 2007, Darmstadt, Germany, (Eds.) 
 P.~Senger et al. 
\bibitem{Fodor} Z.~Fodor, and S.~Katz, J. High Energy Phys. {\bf 0203}, 014 (2002); 
 J. High Energy Phys. {\bf 0404}, 050 (2004). 
\bibitem{GavaiCEP} R.~V.~Gavai, and S.~Gupta, Phys. Rev. D {\bf 71}, 114014 (2005). 
\bibitem{Allton} C.~R.~Allton, S.~Ejiri, S.~J.~Hands, O.~Kaczmarek, F.~Karsch, 
 E.~Laermann, C.~Schmidt, and L.~Scorzato, Phys. Rev. D {\bf 66}, 074507 (2002). 
\bibitem{Koch} V.~Koch, A.~Majumder, and J.~Randrup, Phys. Rev. Lett. {\bf 95}, 182301 (2005). 
\bibitem{Gavaisigns1} R.~V.~Gavai, and S.~Gupta, Phys. Rev. D {\bf 73}, 014004 (2006). 
\bibitem{Gavai} R.~V.~Gavai, S.~Gupta, and P.~Majumdar, Phys. Rev. D {\bf 65}, 054506 (2002). 
\bibitem{Allton1} C.~R.~Allton, S.~Ejiri, S.~J.~Hands, O.~Kaczmarek, F.~Karsch, E.~Laermann, 
  and C.~Schmidt, Phys.~Rev.~D {\bf 68}, 014507 (2003). 
\bibitem{Allton2} C.~R.~Allton, M.~D\"oring, S.~Ejiri, S.~J.~Hands, O.~Kaczmarek, F.~Karsch, 
  E.~Laermann, and K.~Redlich, Phys.~Rev.~D {\bf 71}, 054508 (2005). 
\bibitem{Gavai1} R.~V.~Gavai, and S.~Gupta, Phys. Rev. D {\bf 72}, 054006 (2005). 
\bibitem{Maezawa} Y.~Maezawa, T.~Hatsuda, S.~Aoki, K.~Kanaya, S.~Ejiri, 
 N.~Ishii, N.~Ukita, and T.~Umeda, PoS {\bf LAT2007}, 207 (2007). 
\bibitem{Hietanen} A.~Hietanen, and K.~Rummukainen, PoS {\bf LAT2006}, 137 (2006). 
\bibitem{Bernard} C.~Bernard, T.~Burch, C.~E.~DeTar, J.~Osborn, 
S.~Gottlieb, E.~B.~Gregory, D.~Toussaint, U.~M.~Heller, and R.~Sugar, 
Phys. Rev. D {\bf 71}, 034504 (2005); 
C.~Bernard, T.~Burch, C.~E.~DeTar, S.~Gottlieb, L.~Levkova, U.~M.~Heller, 
 J.~E.~Hetrick, D.~B.~Renner, D.~Toussaint, and R.~Sugar, PoS {\bf LAT2006}, 139 (2006); 
 C.~Bernard, T.~Burch, C.~E.~DeTar, L.~Levkova, S.~Gottlieb, U.~M.~Heller, 
 J.~E.~Hetrick, R.~Sugar, and D.~Toussaint, PoS {\bf LAT2007}, 190 (2007); 
 C.~Bernard, C.~E.~DeTar, L.~Levkova, S.~Gottlieb, U.~M.~Heller, 
 J.~E.~Hetrick, R.~Sugar, and D.~Toussaint, Phys. Rev. D {\bf 77}, 014503 (2008). 
\bibitem{Mukherjee} S.~Mukherjee, Phys. Rev. D {\bf 74}, 054508 (2006). 
\bibitem{Karsch} F.~Karsch, S.~Ejiri, and K.~Redlich, Nucl. Phys. A {\bf 774}, 619 (2006); 
 S.~Ejiri, F.~Karsch, and K.~Redlich, Phys. Lett. B {\bf 633}, 275 (2006). 
\bibitem{Shuryak} J.~Liao, and E.~V.~Shuryak, Phys. Rev. D {\bf 73}, 014509 (2006). 
\bibitem{Shuryak-Zahed} E.~V.~Shuryak, and I.~Zahed,  Phys. Rev. D {\bf 70}, 054507 (2004); 
 B.~A.~Gelman, E.~V.~Shuryak, and I.~Zahed, Phys. Rev. C {\bf 74}, 044908 (2006). 
\bibitem{Peshier-Cassing} A.~Peshier, and W.~Cassing, Phys. Rev. Lett. {\bf 94}, 172301 (2005); 
 W.~Cassing, Nucl. Phys. A {\bf 795}, 70 (2007). 
\bibitem{Sasaki1} C.~Sasaki, B.~Friman, and K.~Redlich, Phys. Rev. D {\bf 75}, 054026 (2007). 
\bibitem{Sasaki2} C.~Sasaki, B.~Friman, and K.~Redlich, Phys. Rev. D {\bf 75}, 074013 (2007). 
\bibitem{Ghosh} S.~K.~Ghosh, T.~K.~Mukherjee, M.~G.~Mustafa, and R.~Ray, Phys. Rev. D {\bf 73}, 114007 (2006); 
 arXiv:0710.2790 [hep-ph]. 
\bibitem{Roessner} S.~R\"o\ss ner, C.~Ratti, and W.~Weise, Phys. Rev. D {\bf 75}, 034007 (2007); 
 Phys. Lett. B {\bf 649}, 57 (2007). 
\bibitem{Bannur} V.~M.~Bannur, Eur. Phys. J. C {\bf 50}, 629 (2007). 
\bibitem{Bluhm_PoS} B.~K\"ampfer, M.~Bluhm, H.~Schade, R.~Schulze, and D.~Seipt, PoS {\bf CPOD2007}, 007 (2007). 
\bibitem{Blaizotchi} J.-P.~Blaizot, I.~Iancu, and A.~Rebhan, Phys. Lett. B {\bf 523}, 143 (2001). 
\bibitem{Petretsky} G.~Boyd, S.~Gupta, and F.~Karsch, Nucl. Phys. B {\bf 385}, 481 (1992); 
 P.~Petreczky, F.~Karsch, E.~Laermann, S.~Stickan, and I.~Wetzorke, Nucl. Phys. Proc. Suppl. {\bf 106}, 513 (2002). 
\bibitem{Karsch-Kitazawa} F.~Karsch, and M.~Kitazawa, Phys. Lett. B {\bf 658}, 45 (2007). 
\bibitem{Peshier1} A.~Peshier, B.~K\"ampfer, O.~P.~Pavlenko, and G.~Soff, Phys. Lett. B {\bf 337}, 235 (1994); 
 Phys. Rev. D {\bf 54}, 2399 (1996). 
\bibitem{Peshier2} A.~Peshier, B.~K\"ampfer, and G.~Soff, Phys. Rev. C {\bf 61}, 045203 (2000); 
 Phys. Rev. D {\bf 66}, 094003 (2002). 
\bibitem{Bluhm_PLB} M.~Bluhm, B.~K\"ampfer, and G.~Soff, Phys. Lett. B {\bf 620}, 131 (2005). 
\bibitem{Bluhm_PRC} M.~Bluhm, B.~K\"ampfer, R.~Schulze, D.~Seipt, and U.~Heinz, Phys. Rev. C {\bf 76}, 034901 (2007). 
\bibitem{Peshier_priv} A.~Peshier, private communication, 2002. 
\bibitem{Seipt_Diploma} D.~Seipt, Diploma Thesis, {\it Quark mass dependence of one-loop self-energies 
 in hot QCD}, Technische Universit\"at Dresden, Germany, May 2007. 
\bibitem{Hatta} Y.~Hatta, and M.~A.~Stephanov, Phys. Rev. Lett. {\bf 91}, 102003 (2003). 
\bibitem{Karschcutoff} F.~Karsch, 
 arXiv:0711.0656 [hep-lat]. 
\bibitem{Karsch1} F.~Karsch, E.~Laermann, and A.~Peikert, Phys. Lett. B {\bf 478}, 447 (2000). 
\bibitem{Karschpriv} F.~Karsch, private communication, January 2006. 
\bibitem{Gavai2} R.~V.~Gavai, and S.~Gupta, Phys. Rev. D {\bf 65}, 094515 (2002); 
 Phys. Rev. D {\bf 67}, 034501 (2003); private communication, March 2006. 
\bibitem{Gavai3} R.~V.~Gavai, and S.~Gupta, Phys. Rev. D {\bf 68}, 034506 (2003). 
\bibitem{Romatschke} A.~Rebhan, and P.~Romatschke, Phys. Rev. D {\bf 68}, 025022 (2003).
\bibitem{PNJL} S.~Mukherjee, M.~G.~Mustafa, and R.~Ray, Phys. Rev. D {\bf 75}, 094015 (2007).
\bibitem{Miao} C.~Miao, and C.~Schmidt, PoS {\bf LAT2007}, 175 (2007). 
\bibitem{Wilczek} K.~Rajagopal, and F.~Wilczek, Phys. Rev. Lett. {\bf 86}, 3492 (2001). 
\bibitem{Bluhm_EPJC} M.~Bluhm, B.~K\"ampfer, R.~Schulze, and D.~Seipt, Eur. Phys. J. C {\bf 49}, 205 (2007). 
\bibitem{Redlich} F.~Karsch, K.~Redlich, and A.~Tawfik, Phys. Lett. B {\bf 571}, 67 (2003).
\bibitem{Blaschke} D.~B.~Blaschke, and K.~A.~Bugaev, Phys. Part. Nucl. Lett. {\bf 2}, 305 (2005).
\bibitem{Weise} M.~Procura, B.~U.~Musch, T.~Wollenweber, T.~R.~Hemmert, and W.~Weise, 
 Phys. Rev. D {\bf 73}, 114510 (2006).
\bibitem{Toneev} A.~S.~Khvorostukin, V.~V.~Skokov, V.~D.~Toneev, and K.~Redlich, 
 Eur. Phys. J. C {\bf 48}, 531 (2006);
 Yu.~B.~Ivanov, A.~S.~Khvorostukhin, E.~E.~Kolomeitsev, V.~V.~Skokov, 
V.~D.~Toneev, and D.~N.~Voskresensky, 
 Phys. Rev. C {\bf 72}, 025804 (2005).
\bibitem{Biro} T.~S.~Biro, P.~Levai, P.~Van, and J.~Zimanyi, Phys. Rev. C {\bf 75}, 034910 (2007). 
\bibitem{Bluhm_QM_POS06}
 M.~Bluhm, and B.~K\"ampfer, PoS {\bf CPOD2006}, 004 (2006);
 B.~K\"ampfer, M.~Bluhm, R.~Schulze, D.~Seipt, and U.~Heinz, Nucl. Phys. A {\bf 774}, 757 (2006).
\bibitem{Baym} G.~Baym, Phys. Rev. {\bf 127}, 1391 (1962). 
\bibitem{Blaizot} J.-P.~Blaizot, E.~Iancu, and A.~Rebhan, 
 Phys. Rev. Lett. {\bf 83}, 2906 (1999); 
 Phys. Lett. B {\bf 470}, 181 (1999); Phys. Rev. D {\bf 63}, 065003 (2001); 
 Phys. Rev. D {\bf 68}, 025011 (2003); 
 and in {\it Quark Gluon Plasma 3}, edited by R.~C.~Hwa and X.~N.~Wang 
 (World Scientific, Singapore, 2004), p. 60. 
\bibitem{Vanderheyden} B.~Vanderheyden, and G.~Baym, J. Stat. Phys. {\bf 93}, 843 (1998); 
 and in {\it Progress in Nonequilibrium Green's functions}, (Ed.) M.~Bonitz (World Scientific, 
 Singapore, 2000). 
\bibitem{leBellac} M.~Le~Bellac, {\it Thermal Field Theory} (Cambridge 
University Press, Cambridge, 1996). 
\bibitem{Bluhm_immu} M.~Bluhm, and B.~K\"ampfer, Phys. Rev. D {\bf 77}, 
034004 (2008). 

\end{thebibliography}
\end{document}